\documentclass[a4paper,11pt]{article}

\usepackage{jheppub}                 
\usepackage{multicol}

\usepackage{paralist}
\usepackage{graphicx}
\usepackage{enumitem}
\usepackage{bm}
\usepackage{color}
\usepackage{amsmath}
\usepackage[utf8]{inputenc}
\usepackage{subfigure}
\usepackage{placeins}
\usepackage[normalem]{ulem}
\usepackage{multirow}
\usepackage{bbold}

\newcommand{\eq}[1]{Eq.~(\ref{#1})}

\newcommand{\beq} {\begin{equation}}
\newcommand{\eeq} {\end{equation}}
\newcommand{\bea} {\begin{eqnarray}}
\newcommand{\eea} {\end{eqnarray}}
\newcommand{\ba} {\begin{eqnarray*}}
\newcommand{\ea} {\end{eqnarray*}}
\newcommand{\GeV} {\,\text{GeV}}
\newcommand{\TeV} {\,\text{TeV}}
\newcommand{\tb}{$\bar{3}$}

\newcommand{\hc}{\mathrm{h.c.}}

\preprint{PSI-PR-19-05, ZU-TH-18/19}

\title{\begin{boldmath}Generic Loop Effects of New Scalars and Fermions in $b\to s\ell^+\ell^-$, $(g-2)_\mu$ and a Vector-like $4^{\rm th}$ Generation\end{boldmath}}

\author{Pere Arnan$^a$,}
\author{Andreas Crivellin$^b$,} 
\author{Marco Fedele$^a$,} 
\author{and Federico Mescia$^a$}

\affiliation{$^a$Departament de F\'isica Quàntica i Astrof\'isica (FQA), Institut de Ciències del Cosmos (ICCUB), Universitat de Barcelona (UB), Spain} 
\affiliation{$^b$Paul Scherrer Institut, CH–5232 Villigen PSI, Switzerland} 

\emailAdd{arnan@ub.edu}
\emailAdd{andreas.crivellin@cern.ch}
\emailAdd{marco.fedele@icc.ub.edu}
\emailAdd{mescia@ub.edu}

\abstract{In this article we investigate in detail the possibility of accounting for the $b\to s\ell^+\ell^-$ and $(g-2)_\mu$ anomalies via loop contributions involving with new scalars and fermions. For this purpose, we first write down the most general Lagrangian which can generate the desired effects and then calculate the generic expressions for all relevant $b\to s$ Wilson coefficients. Here we extend previous analysis by allowing that the new particles can also couple to right-handed Standard Model (SM) fermions as preferred by recent $b\to s\ell^+\ell^-$ data and the anomalous magnetic moment of the muon.
	
In the second part of this article we illustrate this generic approach for a UV complete model in which we supplement the Standard Model by a $4^{\rm th}$ generation of vector-like fermions and a real scalar field. This model allows one to coherently address the observed anomalies in $b\to s\ell^+\ell^-$ transitions and in $a_\mu$ without violating the bounds from other observables (in particular $B_s -\bar B_s$ mixing) or LHC searches. In fact, we find that our global fit to this model, after the recent experimental updates, is very good and prefers couplings to right-handed SM fermions, showing the importance of our generic setup and calculation performed in the first part of the article. 

}

\begin{document} 
\maketitle
\allowdisplaybreaks

\newpage

\section{Introduction}\label{sec:intro}

While no particles beyond the ones of the Standard Model (SM) have been observed at the LHC (so far), $b\to s\ell^+\ell^-$ data show a coherent pattern of deviations from the SM predictions with a significance of more than $4$--$5\,\sigma$~\cite{Capdevila:2017bsm,Altmannshofer:2017yso,DAmico:2017mtc,Hiller:2017bzc,Geng:2017svp,Ciuchini:2017mik,Alok:2017sui,Hurth:2017hxg}. Recently, results of Belle and LHCb presented at Moriond EW 2019~\cite{Aaij:2019wad,Abdesselam:2019wac} confirmed these tensions, even though the significance for the new physics (NP) hypothesis, compared to the SM, did not change notably\footnote{Note that deviations from the SM predictions have been observed in $b\to c\tau\nu$ transitions as well~\cite{Amhis:2016xyh}. However, since these tensions cannot be explained by loop effects, we do not discuss them in this article.}. In fact, including these new measurements, global fits of the Wilson coefficients governing $b \to s\ell^+\ell^-$ transitions~\cite{Alguero:2019ptt,Alok:2019ufo,Ciuchini:2019usw,Datta:2019zca,Aebischer:2019mlg,Kowalska:2019ley} still find that NP scenarios can describe data much better than the SM, even though the preferences between the different scenarios changed with respect to the previous experimental situation. 
\medskip

Concerning concrete NP models giving the desired pattern in the effective theory with a good fit to data, most analyses focused on scenarios in which the required NP effects are generated at tree-level, either by the exchange of $Z^\prime$ vector bosons~\cite{Descotes-Genon:2013wba,Gauld:2013qja,Buras:2013dea,Altmannshofer:2014cfa,Crivellin:2015mga,Crivellin:2015lwa,Niehoff:2015bfa,Sierra:2015fma,Crivellin:2015era,Celis:2015ara,Allanach:2015gkd,Becirevic:2016zri,Celis:2016ayl,Boucenna:2016wpr,Boucenna:2016qad,Megias:2016bde,Altmannshofer:2016jzy,Crivellin:2016ejn,DiChiara:2017cjq,Chiang:2017hlj,Romao:2017qnu,Bian:2017xzg,Falkowski:2018dsl,Guadagnoli:2018ojc} or via leptoquarks~\cite{Gripaios:2014tna,Becirevic:2015asa,Varzielas:2015iva,Alonso:2015sja,Calibbi:2015kma,Belanger:2015nma,Barbieri:2015yvd,Becirevic:2016oho,Becirevic:2016yqi,Sahoo:2016pet,Hiller:2016kry,Cox:2016epl,Crivellin:2017zlb,Cai:2017wry,Dorsner:2017ufx,Buttazzo:2017ixm,Assad:2017iib,DiLuzio:2017vat,Calibbi:2017qbu,Bordone:2017bld,Blanke:2018sro,Marzocca:2018wcf,Bordone:2018nbg,Becirevic:2018afm,Crivellin:2018yvo,deMedeirosVarzielas:2018bcy,DiLuzio:2018zxy,Faber:2018qon,Heeck:2018ntp,Angelescu:2018tyl,Gherardi:2019zil,Cornella:2019hct}. Nonetheless, since the size of the NP contribution required to account for current data is of the order of 20\% compared to the (loop and CKM suppressed) amplitude of the SM, also new loop effects can in principle suffice for an explanation. 
\medskip

In this context, box contributions of heavy new scalars and fermions\footnote{Box contributions of new vectors and fermions were studied in the context of $Z^\prime$ models with vector-like quarks in Ref.~\cite{Bobeth:2016llm}.} (also within multi Higgs doublet models with right-handed neutrinos~\cite{Li:2018rax,Marzo:2019ldg,Crivellin:2019dun}) have been shown to be a viable option~\cite{Gripaios:2015gra,Arnan:2016cpy,Barman:2018jhz,Grinstein:2018fgb}\footnote{Alternatively, models with large couplings to right-handed top quarks can give the desired effect via a $W$-loop~\cite{Becirevic:2017jtw,Kamenik:2017tnu,Camargo-Molina:2018cwu}, as first shown in the EFT context in Ref.~\cite{Aebischer:2015fzz}.}. Furthermore, an explanation of the anomalies in $b\to s\ell^+\ell^-$ via loop effects allows for interesting connections to Dark Matter~\cite{Baek:2017sew,Kawamura:2017ecz,Chiang:2017zkh,Cline:2017qqu,Baek:2018aru,Baek:2019qte,Cerdeno:2019vpd} and typically leads to correlated imprints on other observables like the anomalous magnetic moment of the muon {($a_\mu$)}. However, the effect here is in most models too small since a quite large NP contribution is needed to account from the tantalizing tension between the measurement~\cite{Bennett:2006fi} and the SM prediction of around $3$--$4\,\sigma$. In fact, Ref.~\cite{Arnan:2016cpy} found that it is challenging to account for $\Delta a_\mu$ with TeV scale masses and not too large couplings to muons with a minimal particle content. In general, it has been argued~\cite{Crivellin:2018qmi} that one needs new sources of electroweak symmetry breaking (EWSB) if one aims at a high scale explanation of the anomalous magnetic moment of the muon. In the context of adding new scalars and fermions to the SM this can be achieved for example by a fourth generation of vector-like leptons coupling to the SM Higgs~\cite{delAguila:2008pw,Kannike:2011ng,Joglekar:2012vc,Kearney:2012zi,	Ishiwata:2013gma,Dermisek:2013gta,Kowalska:2017iqv,Calibbi:2018rzv,Crivellin:2018qmi}. 
\medskip

Therefore, we extend in this article the analysis of Ref.~\cite{Arnan:2016cpy} to include the possibility of new sources of EW symmetry breaking within the NP sector. For this purpose, an extension of the field content with respect to the minimal one of Ref.~\cite{Arnan:2016cpy} is necessary, i.e. more than three new fields need to be added to the SM particle content. In doing so, new couplings to right-handed quarks and leptons are introduced which do not only affect $a_\mu$ but also lead to different effects in $b\to s\mu^+\mu^-$ (i.e. lead to solutions other than the purely left-handed $C_9=-C_{10}$ one obtained in Ref.~\cite{Arnan:2016cpy}). In fact, while before Moriond 2019 scenarios with left-handed current were in general preferred, now including right-handed contributions (both in quark and leptonic sectors) can even give a better fit to data~\cite{Alguero:2019ptt,Alok:2019ufo,Ciuchini:2019usw,Aebischer:2019mlg,Kowalska:2019ley}. 
\medskip

A UV complete example of such a setup with new scalars and fermions couplings to left- and right-handed SM fermions is a model with a vector-like $4^{\rm th}$ generation. With respect to Ref.~\cite{Poh:2017tfo,Raby:2017igl}, also aiming at an explanation of the $b\to s\ell^+\ell^-$ anomalies, we add not only a $4^{\rm th}$ generation of leptons but also of quarks~\cite{Branco:1986my,delAguila:1989rq,Langacker:1988ur,delAguila:1997vn} to the SM. However, instead of adding a $Z^\prime$ boson we supplement the model by a neutral scalar to get the desired loop-contributions. Furthermore, one can forbid the dangerous mixing effect between the SM fermions and the new vector-like ones by assigning $U(1)$ changes to the new particles (resembling R-parity in the MSSM).
\medskip

This article is organized as follows: In Sec.~\ref{sec:generic} we define our generic setup, in which new scalars and fermions couple to SM quarks and leptons via Yukawa-like interactions. There, we also provide completely general expressions for the formulae of the relevant Wilson coefficients. We review the corresponding observables together with the current experimental situation in Sec.~\ref{sec:Exp}. Our generic approach of Sec.~\ref{sec:generic} is then applied to a specific UV complete model in Sec.~\ref{sec:4thgen}, which contains a vector-like fourth generation of fermions and a neutral scalar. We study the phenomenology of this model in detail before we conclude in Sec.~\ref{sec:conclusions}.
\medskip


\section{Generic Setup and Wilson Coefficients}
\label{sec:generic}

\begin{figure}[!t]
	\centering
	\includegraphics[scale=0.56]{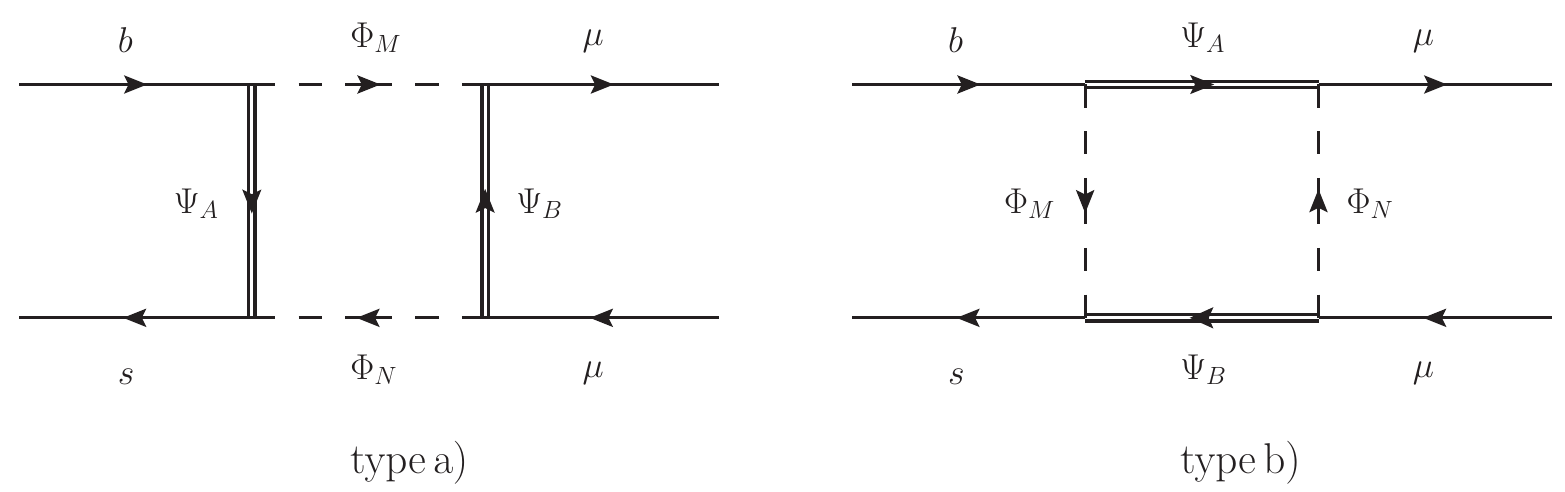}
	\caption{Box diagrams contributing to $b\to s\mu^+\mu^-$ transitions. The diagram on the left is generated in models in which the fermions couples only to SM quarks or only to SM leptons, which corresponds to type $a)$. The diagram on the right refers to models with scalars connecting $b$ to $s$ and $\mu$ to $\mu$, i.e. type $b)$.}
	\label{fig:diagbsmumu}
\end{figure}

In this section we define our generic setup and calculate completely general 1-loop expressions for contributions to $b\to s$ processes and the anomalous magnetic moment of the muon. 

As outlined in the introduction, in the spirit of Refs.~\cite{Gripaios:2015gra,Arnan:2016cpy} we add to the SM particle content a NP sector with vector-like fermions $\Psi_{A}$ and new scalars $\Phi_{M}$  such that $b \to s \mu^+\mu^-$ transitions can be generated via box diagrams, as depicted in Fig.~\ref{fig:diagbsmumu}. In this respect, we generalize the previous analysis of Ref.~\cite{Arnan:2016cpy} by including in addition couplings of new particles to $SU(2)$ singlet SM fermions. Moreover, we do not impose limitations on the number of fields added to the SM and allow for couplings of the new sector to the SM Higgs. 

In order to generate box diagrams as the ones shown in Fig.~\ref{fig:diagbsmumu} it is necessary that either the scalars $\Phi_{M,N}$ or the fermions $\Psi_{A,B}$ couple both to quarks and leptons, corresponding to case $a)$ and $b)$, respectively. This means that in diagrams of type $a)$ the amplitudes (before using any Fierz identities) have the structure $(\bar s \Gamma b) (\bar \mu \Gamma \mu)$, while in type $b)$ amplitudes of the form $(\bar \mu \Gamma b) (\bar s \Gamma \mu)$ are generated. Here, $\Gamma$ denotes an arbitrary Dirac structure. Since semi-leptonic operators are commonly given in the form $(\bar s \Gamma b) (\bar \mu \Gamma \mu)$, Fierz identities must be used in case $b)$ in order to transform the expressions to this standard basis. We give the relevant Fierz identities in Appendix~\ref{app:fierz}.

The Yukawa-like couplings of new scalars $\Phi_M$ and fermions $\Psi_A$ to bottom/strange quarks and muons can be parameterized completely generically (below the EWSB scale) by the Lagrangian
\begin{eqnarray} \label{eq:L_bsmumu}
\mathcal{L}_{\rm int} = & \bigg[ & \bar \Psi _A\left( L_{AM}^b{P_L}{b } + L_{AM}^s{P_L}{s } + L_{AM}^\mu {P_L}\mu
\right){\Phi _M}  \nonumber\\
&& + \bar \Psi _A\left( {R_{AM}^b{P_R}{b } + R_{AM}^s{P_R}{s } + R_{AM}^\mu {P_R}\mu } \right){\Phi _M} \bigg] + \hc\, .
\end{eqnarray}
Here $\Psi_A$ and $\Phi_M$ have to be understood as generic lists containing in principle an arbitrary number of fields, meaning that $A$ and $M$ also include implicitly $SU(2)$ and color indices. Therefore, the couplings $L_{AM}^{s,b}$ and $R_{AM}^{s,b}$ are generic matrices in ($A$-$M$) space with the restriction that $U(1)_{\rm EM}$ and $SU(3)$ are respected\footnote{Here we only consider coupling to muons in order to explain the anomalies in $b\to s\ell^+\ell^-$. The reason for this is that in our setup sizable couplings to electrons would in general generate effects in $\mu\to e\gamma$, which would contradict experimental bounds~\cite{TheMEG:2016wtm} by orders of magnitude.}. 

This Lagrangian will not only affect $b\to s\mu^+\mu^-$ transitions but also unavoidably generate effects in $B_s - \bar{B}_s$ mixing, $b \to s \gamma$ decays, the anomalous magnetic moment of the muon $a_\mu$ as well as $Z$ couplings and decays to SM fermions. Furthermore, $b\to s\nu\bar\nu$ processes and $D_0-\bar D_0$ mixing can give relevant constraints once $SU(2)$ invariance at the NP scale is imposed.  Therefore, all these processes have to be taken into account in a complete phenomenological analysis. In order to perform such an analysis, the Wilson coefficients of the relevant effective Hamiltonian must be known. We will calculate them in the following subsections.
\medskip


\boldmath
\subsection{\texorpdfstring{$b\to s \mu^+\mu^-$}{btosmu+mu-} and \texorpdfstring{$b \to s \gamma$}{btosgamma} Transitions}\label{sec:bs_gen}
\unboldmath

The dimension-6 operators governing $b\to s \mu^+\mu^-$ and $b \to s \gamma$ transitions are contained in the effective Hamiltonian:
\begin{equation}
\label{eq:heffbsmumu}
\mathcal{H}_\mathrm{eff}^{\ell\ell} = - \frac{4 G_F}{\sqrt{2}}V_{tb}V_{ts}^\ast \sum\limits_{\substack{i}}\Bigg{(}C_i(\mu)\mathcal{O}_i(\mu)+C_i^\prime(\mu)\mathcal{O}_i^\prime(\mu)\Bigg{)} + \hc\,,
\end{equation}
where
\begin{alignat}{3}
\mathcal{O}_7 &= \frac{e}{16\pi^2} m_b \bar{s}\sigma^{\mu\nu} P_R b F_{\mu \nu}\,,\qquad\qquad
&& \mathcal{O}_8 &&=\frac{g_s}{16\pi^2} m_b \bar{s}_\alpha\sigma^{\mu\nu} P_R T^a_{\alpha\beta}b_\beta G_{\mu \nu}^a\,,  \nonumber\\
\mathcal{O}_9 &= \frac{\alpha _{\text{EM}}}{4\pi}(\bar{s}\gamma_\mu P_L b)(\bar{\mu}\gamma^\mu \mu)\,, \qquad\qquad  
&& \mathcal{O}_{10} &&=\frac{\alpha _{\text{EM}}}{4\pi}(\bar{s}\gamma_\mu P_L b)(\bar{\mu}\gamma^\mu \gamma_5 \mu)\,,  \nonumber\\
\mathcal{O}_S &=\frac{\alpha _{\text{EM}}}{4\pi}(\bar{s} P_R b)(\bar{\mu} \mu)\, ,\qquad\qquad 
&& \mathcal{O}_P &&=\frac{\alpha _{\text{EM}}}{4\pi}(\bar{s} P_R b)(\bar{\mu}\gamma_5 \mu)\,,   \nonumber\\
\mathcal{O}_{T} &=\frac{\alpha _{\text{EM}}}{4\pi}(\bar{s}\sigma_{\mu\nu} b)(\bar{\mu}\sigma^{\mu\nu} P_R \mu)\,,
\end{alignat}
with $e$ being the electron charge, $\alpha_{\text{EM}}$ the fine structure constant and $g_s$ the $SU(3)$ gauge coupling. The primed operators are obtained by interchanging $L$ and $R$. NP contributions from box diagrams will generate effects in $C^{(\prime)}_{9,10}$, $C^{(\prime)}_{S,P}$ and $O^{(\prime)}_{T}$, while on-shell photon (gluon) penguins generate $C^{(\prime)}_{7(8)}$ and $C^{(\prime)}_{9}$, and $Z$-penguins $C^{(\prime)}_{9,10}$.
\medskip
\begin{table}[!t]
\centering
\begin{tabular}{c | cccc | cccc | c}
 \multirow{2}{*}{ $SU(3)$}  & \multicolumn{4}{c |}{$b \to s\ell\bar\ell$\; type $a)$} & \multicolumn{4}{c |}{$b \to s\ell\bar\ell$\; type $b)$} &  \multirow{2}{*}{$\chi$}\\
 \cline{2-9} 
 & $\Psi_A$ & $\Psi_B$ & $\Phi_M$ & $\Phi_N$ &  $\Psi_A$ & $\Psi_B$ & $\Phi_M$ & $\Phi_N$ \\
\hline
I   & 3   & 1   & 1   & 1       & 1   & 1   & \tb & 1    &   1 \\  
II  & 1   & \tb & \tb & \tb     & 3   & 3   & 1   & 3    &   1 \\ 
III & 3   & 8   & 8   & 8       & 8   & 8   & \tb & 8    &   4/3 \\
IV   & 8   & \tb & \tb & \tb     & 3   & 3   & 8   & 3    &   4/3 \\ 
V   & \tb & 3   & 3   & 3       & \tb & \tb & 3   & \tb  &   2
\end{tabular}
\caption{Table of the possible $SU(3)$ representations that can give an effect in $b \to s\ell^+\ell^-$ or $b \to s\nu\nu$ transitions via box diagrams. $\chi$ denotes the resulting group factor appearing in Eqs.~\eqref{eq:C9_gen}-\eqref{eq:CT_gen} which also enters in $b \to s\nu\nu$ transitions. \label{tab:SU3_bsmumu}}
\end{table} 

The box diagrams in Fig.~\ref{fig:diagbsmumu}, result in the following Wilson coefficients (here and in the remainder of the section, an implicit sum over all NP particles, i.e. $A,B,M,N$, is understood):
\begin{eqnarray}
C_9^{\text{box},\, a)} &=& - {\cal N}
\frac{\chi L^{s*}_{AN}L^b_{AM}}{{32\pi \alpha_{\text{EM}}  m_{\Phi_M}^2}}
\left[ L^{\mu *}_{BM}L^\mu_{BN} + R^{\mu *}_{BM}R^\mu_{BN}\right]
F(x_{AM},x_{BM},x_{NM})\,, \nonumber\\\label{eq:C9_gen}
C_9^{{\text{box},\, b)}} &=& -{\cal N}
\frac{\chi L^{s*}_{BM}L^b_{AM}}{{32\pi \alpha_{\text{EM}} m_{\Phi_M}^2}}
\bigg[ L^{\mu *}_{AN}L^\mu_{BN} F(x_{AM},x_{BM},x_{NM}) \nonumber\\
&& \qquad\qquad\qquad\qquad   
- R^{\mu *}_{AN}R^\mu_{BN} \frac{m_{\Psi_A} m_{\Psi_B}}{m_{\Phi_M}^2}G(x_{AM},x_{BM},x_{NM}) \bigg]\,,
\\
C_{10}^{{\text{box},\, a)}} &=& {\cal N}
\frac{\chi L^{s*}_{AN}L^b_{AM}}{{32\pi \alpha_{\text{EM}} m_{\Phi_M}^2}}
\left[ L^{\mu *}_{BM}L^\mu_{BN} - R^{\mu *}_{BM}R^\mu_{BN}\right]
F(x_{AM},x_{BM},x_{NM})\,, \nonumber\\\label{eq:C10_gen}
C_{10}^{{\text{box},\, b)}} &=& {\cal N}
\frac{\chi L^{s*}_{BM}L^b_{AM}}{{32\pi \alpha_{\text{EM}} m_{\Phi_M}^2}}
\bigg[ L^{\mu *}_{AN}L^\mu_{BN} F(x_{AM},x_{BM},x_{NM}) \nonumber\\
&& \qquad\qquad\qquad\quad + R^{\mu *}_{AN}R^\mu_{BN} \frac{m_{\Psi_A} m_{\Psi_B}}{m_{\Phi_M}^2} G(x_{AM},x_{BM},x_{NM}) \bigg]\,,
\\
C_S^{{\text{box},\, a)}} &=& - {\cal N}
\frac{\chi L^{s*}_{AN}R^b_{AM}}{{16\pi \alpha_{\text{EM}} m_{\Phi_M}^2}}
\left[ R^{\mu *}_{BM}L^\mu_{BN} + L^{\mu *}_{BM}R^\mu_{BN}\right]
\frac{m_{\Psi_A} m_{\Psi_B}}{m_{\Phi_M}^2} G(x_{AM},x_{BM},x_{NM})\,, \nonumber\\\label{eq:CS_gen}
C_S^{{\text{box},\, b)}} &=& {\cal N}
\frac{\chi L^{s*}_{BM}R^b_{AM}}{{16\pi \alpha_{\text{EM}} m_{\Phi_M}^2}}
\bigg[ R^{\mu *}_{AN}L^\mu_{BN}   F(x_{AM},x_{BM},x_{NM}) \nonumber\\
&& \qquad\qquad\qquad\quad 
+ L^{\mu *}_{AN}R^\mu_{BN}\frac{m_{\Psi_A} m_{\Psi_B}}{2m_{\Phi_M}^2} G(x_{AM},x_{BM},x_{NM}) \bigg]\,,
\\
C_P^{{\text{box},\, a)}} &=& {\cal N}
\frac{\chi L^{s*}_{AN}R^b_{AM}}{{16\pi \alpha_{\text{EM}} m_{\Phi_M}^2}}
\left[ R^{\mu *}_{BM}L^\mu_{BN} - L^{\mu *}_{BM}R^\mu_{BN} \right]
\frac{m_{\Psi_A} m_{\Psi_B}}{m_{\Phi_M}^2} G(x_{AM},x_{BM},x_{NM})\,, \nonumber\\\label{eq:CP_gen}
C_P^{{\text{box},\, b)}} &=& - {\cal N}
\frac{\chi L^{s*}_{BM}R^b_{AM}}{{16\pi \alpha_{\text{EM}} m_{\Phi_M}^2}}
\bigg[ R^{\mu *}_{AN}L^\mu_{BN} F(x_{AM},x_{BM},x_{NM}) \nonumber\\
&& \qquad\qquad\qquad\quad 
- L^{\mu *}_{AN}R^\mu_{BN}\frac{m_{\Psi_A} m_{\Psi_B}}{2m_{\Phi_M}^2} G(x_{AM},x_{BM},x_{NM}) \bigg]\,,
\\
C_{T}^{\text{box},\, b)} &=& -{\cal N}
\frac{\chi
L^{s*}_{BM}R^b_{AM} L^{\mu *}_{AN}R^\mu_{BN}
}{{16\pi \alpha_{\text{EM}} m_{\Phi_M}^2}}
\frac{m_{\Psi_A} m_{\Psi_B}}{m_{\Phi_M}^2}  G(x_{AM},x_{BM},x_{NM})\,, \label{eq:CT_gen} 
\end{eqnarray}
\begin{table}[!t]
	\centering
	\begin{tabular}{c | cc | ccc}
		$SU(3)$ & $\Psi_A$ & $\Phi_M$ & $\chi_\gamma$ & $\chi_g$ & $\tilde\chi_g$ \\
		\hline
		I   & 3   & 1    & 1   &  1   &  0 \\  
		II    & 1   & \tb  & 1   &  0   &  1 \\  
		III   & 3   & 8    & 4/3 & -1/6 &  3/2 \\  
		IV   & 8   & \tb  & 4/3 &  3/2 & -1/6 \\ 
		V  & \tb & 3    & 2   &  -1  & 1
	\end{tabular}
	\caption{Table of the different $SU(3)$ representations that can give non-zero effects via photon- and gluon-penguin diagrams to $b \to s\mu^+\mu^-$ transitions. $\chi_\gamma$ denotes the resulting group factor for the former contribution, while $\chi_g$ and $\tilde\chi_g$ represent the resulting group factors for the latter.
		\label{tab:SU3_bsg}}
\end{table} 
\begin{eqnarray}
C_{9,S,T}^{\prime \text{box}\,} = C_{9,S,T}^{\text{box}\,} \left(L\leftrightarrow R \right)\,,
\qquad\qquad 
C_{10,P}^{\prime \text{box}\,} = -C_{10,P}^{\text{box}\,} \left(L\leftrightarrow R \right)\,, \label{eq:C910SPp_gen}
\end{eqnarray}
where we have defined 
\begin{equation}
x_{AM} \equiv (m_{\Psi_A}/m_{\Phi_M})^2, \qquad
x_{BM} \equiv (m_{\Psi_B}/m_{\Phi_M})^2, \qquad x_{NM} \equiv (m_{\Phi_N}/m_{\Phi_M})^2\,,
\end{equation} 
and
\begin{equation}\label{eq:Ninv}
{\cal N}^{-1} = \dfrac{4G_F}{\sqrt{2}}V_{tb} V_{ts}^*\,.
\end{equation}
In the equations above, the labels $A$, $B$, $M$ and $N$ denote the particle (in case of several representations) and also include $SU(2)$ components, while the sum over $SU(3)$ indices is encoded in the group factors $\chi$. The dimensionless loop functions $F$ and $G$ are defined in Appendix~\ref{app:loopfunctions}.

Such box contributions are only possible if both color and electric charge are conserved. While the Wilson coefficients of $b\to s\ell\ell$ operators are insensitive to the electric charge of the particle in the box, concerning $SU(3)$, the different possible representations of the new particles lead to distinct group factors $\chi$ in Eqs.~\eqref{eq:C9_gen}-\eqref{eq:CT_gen}. These group factors are different for type $a)$ and $b)$ and are given for all the possible representations in Tab.~\ref{tab:SU3_bsmumu}. Furthermore, crossed box diagrams can be constructed in some particular cases. We give the corresponding expressions for such in Appendix~\ref{app:crossedbsmumu} for the real scalar (or Majorana fermion) case and in Appendix~\ref{app:Crossedbsmumucomplex} for the crossed diagrams arising with complex scalars.
\medskip

\begin{figure}[!t]
	\centering
	\includegraphics[scale=0.5]{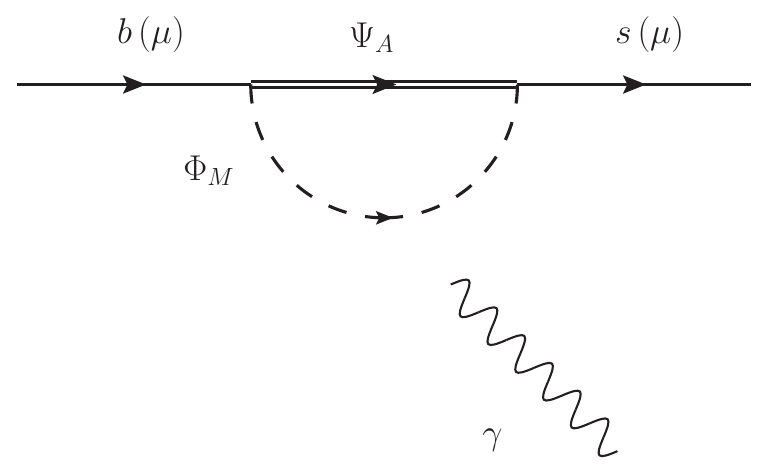}
	\caption{Photon-penguin diagrams contributing to $b\to s \gamma$ transitions and $a_\mu$.}
	\label{fig:photonPenguin}
\end{figure}

On-shell photon penguins diagrams in Fig.~\ref{fig:photonPenguin} affect $C_7^{(\prime)}$ while off-shell ones enter $C_9^{\gamma(\prime)}$:
\bea
C_7^{} &=& {\cal N} \label{eq:C7_gen}
\dfrac{\chi_\gamma L^b_{AM}}{2m_{\Phi_M}^2}\,
\left[
L^{s*}_{AM} \left(Q_{\Phi_M} \widetilde F_7\left(x_{AM}\right) - Q_{\Psi_A} F_7\left(x_{AM}\right)\right) \right. \nonumber\\
&& \qquad\qquad\quad \left. + R^{s*}_{AM} \dfrac{4 m_{\Psi_A}}{m_b} \left(Q_{\Phi_M} \widetilde G_7\left(x_{AM}\right) - Q_{\Psi_A} G_7\left(x_{AM}\right)\right)
\right]\,,
\\
C_9^{\gamma} &=& {\cal N}
\frac{\chi_\gamma L^{s*}_{AM} L^b_{AM}}{{2m_{\Phi_M}^2}} \left[ 
{{Q_{\Phi_M} }{\widetilde F_9}\!\left(x_{AM} \right) - Q_{\Psi_A} {\widetilde G_9}\!\left(x_{AM}\right)} \right]\,,
\label{eq:C9g_gen} \\
C_7^{\prime}&=&C_7\left(L\leftrightarrow R \right)\,, \qquad\qquad
C_9^{\gamma\prime}=C_9^\gamma\left(L\leftrightarrow R \right)\,, \label{eq:C79pg_gen}
\eea
where $m_b$ is the $b$ quark mass. $Q_{\Phi_M}$ and $Q_{\Psi_A}$ are the electric charges of the NP fields ${\Phi_M}$ and ${\Psi_A}$, respectively. The conservation of electric charge imposes that $Q_{\Phi_M} + Q_{\Psi_A} = Q_d \equiv -1/3$. The color factors $\chi_\gamma$, which depend on the $SU(3)$ representations of the new particles in the loop, are given in Tab.~\ref{tab:SU3_bsg}. The loop functions are defined in Appendix~\ref{app:loopfunctions}.
Note that the terms proportional to ${\widetilde F_7}$, ${\widetilde G_7}$ and ${\widetilde F_9}$ in Eqs.~(\ref{eq:C7_gen})-(\ref{eq:C79pg_gen}) stem from the diagram where the photon couples to the scalar $\Phi_{M}$, while the terms proportional to $F_7$, $G_7$ and ${\widetilde G}_9$ stem from the diagram where the photon couples to the fermion $\Psi_{A}$. 
\medskip

Similarly, the gluon-penguin generates
\bea
C_8^{} &=& {\cal N} 
\dfrac{ L^b_{AM}}{2m_{\Phi_M}^2}\,
\left[
L^{s*}_{AM} \left(\chi_g \widetilde F_7\left(x_{AM}\right) - \tilde{\chi}_{g} F_7\left(x_{AM}\right)\right) \right. \nonumber\\
&& \qquad\qquad\quad \left. + R^{s*}_{AM} \dfrac{4 m_{\Psi_A}}{m_b} \left(\chi_g \widetilde G_7\left(x_{AM}\right) - \tilde{\chi}_{g} G_7\left(x_{AM}\right)\right)
\right] \,,\\
C_8^{\prime}&=&C_8\left(L\leftrightarrow R \right)\,, \label{eq:C8g_gen}
\eea
where the color factors $\chi_g$ and $\tilde{\chi}_{g}$ for the different possible $SU(3)$ representations are given in Tab.~\ref{tab:SU3_bsg}. 

\medskip
The contribution of $Z$-penguins to $C_{9,10}^{(\prime)}$ is given in Sec.~\ref{sec:Zmumu} together with a discussion of Z decays.


\boldmath
\subsection{\texorpdfstring{$b\to s\nu\bar{\nu}$}{btosnubarnu} }\label{sec:BKnunu}
\unboldmath

As stated at the beginning of this section, $b\to s\nu_\mu\bar\nu_\mu$ processes have to be taken into account once $SU(2)$ invariance at the NP scale is imposed. This implies that, in the generic description in \eq{eq:L_bsmumu}, one has to replace the left-handed muon fields with neutrinos. The box diagrams generating $b\to s\nu_\mu\bar\nu_\mu$ are therefore obtained from Fig.~\ref{fig:diagbsmumu} by replacing muons with neutrinos.
\medskip

The effective Hamiltonian describing this process reads (following the conventions of Ref.~\cite{Buras:2014fpa})
\beq
{{\cal H}_{\rm eff}^{\nu_\mu} } = 
 - \frac{{4{G_F}}}{{\sqrt 2 }}{V_{tb}}V_{ts}^*\;
 \left(
 C_L{\mathcal{O}_L}
+
 C_R{\mathcal{O}_R}
 \right) +\text{h.c.}\,,
\label{eq:Heff_bsnunu}
\eeq
where
\begin{equation}
\mathcal{O}_{L(R)} = \frac{\alpha_\mathrm{EM} }{{4\pi }} [\bar s{\gamma ^\mu }{P_{L(R)}}b][{{\bar \nu }_\mu}{\gamma _\mu }\left( {1 - {\gamma ^5}} \right){\nu _\mu}]\,.
\end{equation}
The resulting WCs are:
\bea
C_L^{a)} & = & - {\cal N}
\frac{\chi  L^{s*}_{AN} L^b_{AM} L^{\mu *}_{BM} L^\mu_{BN}}{{32\pi \alpha_{\text{EM}} m_{\Phi_M}^2}} 
F(x_{AM},x_{BM},x_{NM})\,,
\nonumber\\
C_L^{b)} & = & {\cal N}
\frac{\chi L^{s*}_{BN} L^b_{AM} L^{\mu *}_{AN} L^\mu_{BN}}{{32\pi \alpha_{\text{EM}} m_{\Phi_M}^2}}
F(x_{AM},x_{BM},x_{NM})\,,
\label{eq:CL22_gen}\\
C_R^{a)}& = & - {\cal N}
\frac{\chi R^{s*}_{AN} R^b_{AM} L^{\mu *}_{BM} L^\mu_{BN}}{{32\pi \alpha_{\text{EM}}  m_{\Phi_M}^2}}
F(x_{AM},x_{BM},x_{NM})\,,
\nonumber\\
C_R^{b)} & = & - {\cal N}
\frac{\chi R^{s*}_{BN} R^b_{AM} L^{\mu *}_{AN} L^\mu_{BN}}{{32\pi \alpha_{\text{EM}}  m_{\Phi_M}^2}}
\frac{m_{\Psi_A} m_{\Psi_B}}{m_{\Phi_M}^2} G(x_{AM},x_{BM},x_{NM})\,,
\label{eq:CR22_gen}
\eea
where the normalization factor $\cal N$ has been introduced in Eq.~(\ref{eq:Ninv}), and the loop functions $F(x,y,z)$ and $G(x,y,z)$ are defined in Appendix~\ref{app:loopfunctions}. The colour factor $\chi$ is the same as for $b\to s\mu^+\mu^-$ transitions and is given in Tab.~\ref{tab:SU3_bsmumu} for the different representations.
\medskip


\boldmath
\subsection{\texorpdfstring{$\Delta B=\Delta S=2$}{DeltaB=DeltaS=2} Processes}
\unboldmath

\begin{figure}[!t]
\centering
\includegraphics[scale=0.5]{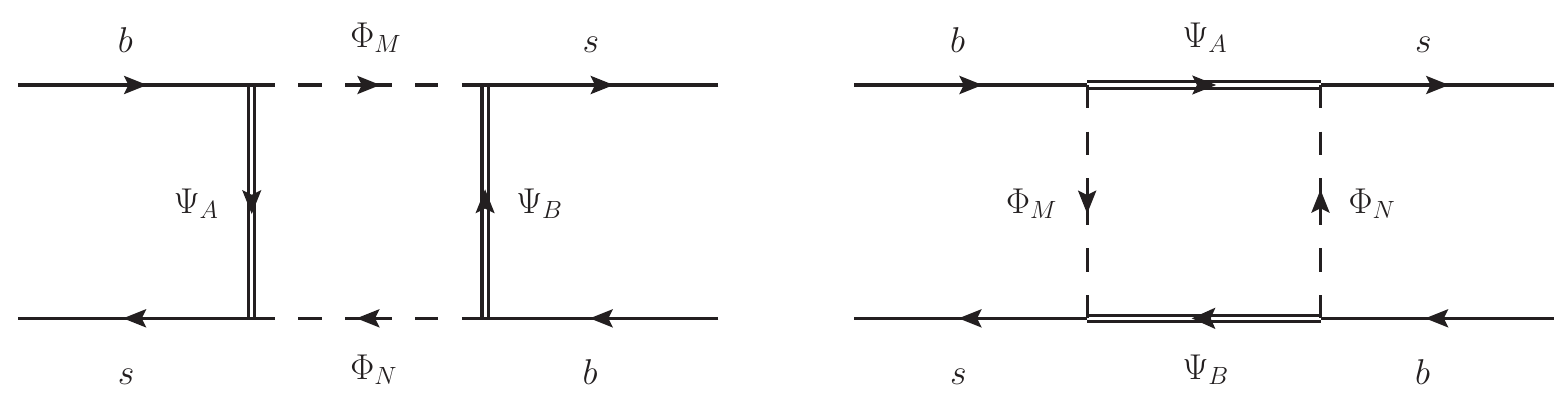}
\caption{Box diagrams contributing to $B_s - {\bar B}_s$ mixing. Both diagrams arise independently of the nature of the mediator involved in $b \to s \mu^+\mu^-$ transitions.}
\label{fig:diagBBmix}
\end{figure}

The presence of $L_{AM}^{b,s}$ and $R_{AM}^{b,s}$ implies NP contributions to the $B_s-\bar{B}_s$ mixing which, using the conventions of Refs.~\cite{Gabbiani:1996hi,Bagger:1997gg}, is governed by 
\begin{equation}\label{eq:heffBBbar}
\mathcal{H}_{\rm eff}^{B_s\bar{B}_s}=C_i \sum_{i=1}^5 \mathcal{O}_i + \tilde{C}_i \sum_{i=1}^3 \widetilde{\mathcal{O}}_i+\text{h.c.} \,,
\end{equation}
with
\begin{equation}
\begin{split}
\mathcal{O}_1&= ({\bar s}_{\alpha} \gamma^{\mu} P_{L} b_{\alpha})\, ({\bar s}_{\beta} \gamma^{\mu} P_{L} b_{\beta})    \,, \qquad  
\widetilde{\mathcal{O}}_1= ({\bar s}_{\alpha} \gamma^{\mu} P_{R} b_{\alpha})\, ({\bar s}_{\beta} \gamma^{\mu} P_{R} b_{\beta})  \,, \\
\mathcal{O}_2&=({\bar s}_{\alpha}  P_{L} b_{\alpha})\, ({\bar s}_{\beta}  P_{L} b_{\beta})  \,,  \qquad \qquad
\widetilde{\mathcal{O}}_2=({\bar s}_{\alpha}  P_{R} b_{\alpha})\, ({\bar s}_{\beta}  P_{R} b_{\beta})  \,, \\
\mathcal{O}_3&=({\bar s}_{\alpha} P_{L} b_{\beta})\, ({\bar s}_{\beta}  P_{L} b_{\alpha})\,, \qquad \qquad
\widetilde{\mathcal{O}}_3=({\bar s}_{\alpha} P_{R} b_{\beta})\, ({\bar s}_{\beta}  P_{R} b_{\alpha})\,, \\
\mathcal{O}_4&=({\bar s}_{\alpha}  P_{L} b_{\alpha})\, ({\bar s}_{\beta}  P_{R} b_{\beta})\,,\\
\mathcal{O}_5&=({\bar s}_{\alpha} P_{L} b_{\beta})\, ({\bar s}_{\beta}  P_{R} b_{\alpha})\,.
\end{split}
\label{eq:OPBB}
\end{equation}

\begin{table}[!t]
\centering
\begin{tabular}{c | cccc | cc}
$SU(3)$ & $\Psi_A$ & $\Psi_B$ & $\Phi_M$ & $\Phi_N$ & $\chi_{BB}$ & $\tilde{\chi}_{BB}$ \\
\hline
I   & 3     & 3     & 1     & 1     &  1    &  0\\  
II  & 1     & 1     & \tb   & \tb   &  0    &  1\\ 
III & 3     & 3     & 8     & 8     &  1/36 &  7/12 \\
IV  & 8     & 8     & \tb   & \tb   &  7/12 &  1/36 \\ 
V   & 3     & 3     & (1,8) & (8,1) & -1/6  &  1/2  \\  
VI  & (1,8) & (8,1) & \tb   & \tb   &  1/2  & -1/6  \\  
VII & \tb   & \tb   & 3     & 3     &  1    &  1  
\end{tabular}
\caption{Table of the different $SU(3)$ representations that can give a non-zero effect via box diagrams to $B_s-\bar{B}_s$ mixing. $\chi_{BB}$ and $\tilde{\chi}_{BB}$ denote the resulting group factors.
\label{tab:SU3_bsbs}}
\end{table}  

The box diagrams contributing to these above operators are shown in Fig.~\ref{fig:diagBBmix}. Using the Lagrangian from Eq.~(\ref{eq:L_bsmumu}), one obtains the following results for the coefficients:
\bea
C_1 &=& (\chi_{BB} + \tilde \chi_{BB}) \label{eq:CBB_gen_beg}
\frac{L^{s*}_{AN}L^b_{AM} L^{s*}_{BM}L^b_{BN}}{128 \pi^2 m_{\Phi_M}^2} 
F\hspace{-2.8pt}\left(x_{AM},x_{BM},x_{NM} \right) \,,	
\\
C_2 &=& \chi_{BB}
\frac{R^{s*}_{AN}L^b_{AM}R^{s*}_{BM}L^b_{BN}}{64 \pi^2 m_{\Phi_M}^2} 
\frac{m_{\Psi_A} m_{\Psi_B}}{m_{\Phi_M}^2} G\hspace{-2.8pt}\left(x_{AM},x_{BM},x_{NM} \right) \,,
\\
C_3 &=& \tilde \chi_{BB}
\frac{R^{s*}_{AN}L^b_{AM}R^{s*}_{BM}L^b_{BN}}{64 \pi^2 m_{\Phi_M}^2} 
\frac{m_{\Psi_A} m_{\Psi_B}}{m_{\Phi_M}^2} G\hspace{-2.8pt}\left(x_{AM},x_{BM},x_{NM} \right) \,,
\\
C_4 &=& \chi_{BB}
\frac{R^{s*}_{AN} L^b_{AM} L^{s*}_{BM} R^b_{BN}}{32 \pi^2 m_{\Phi_M}^2} 
\frac{m_{\Psi_A} m_{\Psi_B}}{m_{\Phi_M}^2}G\hspace{-2.8pt}\left(x_{AM},x_{BM},x_{NM} \right)
\nonumber\\
&& - \tilde \chi_{BB}
\frac{R^{s*}_{AN} R^b_{AM} L^{s*}_{BM} L^b_{BN}}{32 \pi^2 m_{\Phi_M}^2} 
F\hspace{-2.8pt}\left(x_{AM},x_{BM},x_{NM} \right) \,,
\\
C_5 &=&  \tilde \chi_{BB}
\frac{R^{s*}_{AN} L^b_{AM} L^{s*}_{BM} R^b_{BN}}{32 \pi^2 m_{\Phi_M}^2} 
\frac{m_{\Psi_A} m_{\Psi_B}}{m_{\Phi_M}^2}G\hspace{-2.8pt}\left(x_{AM},x_{BM},x_{NM} \right)
 \nonumber \\
&& - \chi_{BB}
\frac{R^{s*}_{AN} R^b_{AM} L^{s*}_{BM} L^b_{BN}}{32 \pi^2 m_{\Phi_M}^2} 
F\hspace{-2.8pt}\left(x_{AM},x_{BM},x_{NM} \right) \,,
\\
\widetilde C_{1,2,3} &=& C_{1,2,3} \left(L \leftrightarrow R \right)\,.
  \label{eq:CBB_gen_end}
\eea
The loop functions $F(x,y,z)$ and $G(x,y,z)$ are defined in Appendix~\ref{app:loopfunctions} and the colour factors $\chi_{BB}$ and $\tilde{\chi}_{BB}$ are given in Tab.~\ref{tab:SU3_bsbs} for the different allowed representations. Again, in the presence of Majorana fermions or real scalars crossed diagrams can be constructed and the resulting expressions are given in Appendix~\ref{app:crossedBs}.
\medskip

\subsection{\texorpdfstring{$D_0 - \bar{D}_0$}{D0-barD0} Mixing}

NP contributions to the $D_0 - \bar{D}_0$ mixing can be obtained in complete generality (at the low scale) from Eqs.~\eqref{eq:CBB_gen_beg}-\eqref{eq:CBB_gen_end} by making the substitutions $s\to u$, $b\to c$, introducing couplings $L^{u,c}_{AM}$ and $R^{u,c}_{AM}$ of new scalars and fermions to up-quarks in straightforward extension of \eq{eq:L_bsmumu}.
\medskip

In the context a UV complete model, $SU(2)$ invariance imposes at the high scale that couplings to left-handed up-type quarks are related to the couplings to left-handed down-type quarks via CKM rotations. Therefore, working in the down-basis, the ``minimal" effect generated in $D_0 - \bar{D}_0$ is induced by the couplings
\begin{equation}
L^{u}_{AM}= V_{us}^* L^{s}_{AM} + V_{ub}^* L^{b}_{AM}\,,\qquad L^{c}_{AM}= V_{cs}^* L^{s}_{AM} + V_{cb}^* L^{b}_{AM}\,.    
\end{equation}


\subsection{Anomalous Magnetic Moment of the Muon}\label{sec:gm2_gen}

\begin{table}[!t]
\centering
\begin{tabular}{c | cc | c}
$SU(3)$ & $\Psi_A$ & $\Phi_M$ & $\chi_{a_\mu}$  \\
\hline
I   & 1       & 1       & 1 \\  
II   & (3,\tb) & (3,\tb) & 3 \\  
III   & 8       & 8       & 8 \\
\end{tabular}
\caption{Table of the different $SU(3)$ representations that can give a non-zero effect to $a_\mu$. $\chi_{a_\mu}$ denotes the resulting group factor.
\label{tab:SU3_gm2}}
\end{table}

The anomalous magnetic moment of the muon ($a_\mu \equiv (g-2)_\mu/2$) and its electric dipole moments ($d_\mu$) we find from the diagrams in Fig.~\ref{fig:photonPenguin}
\bea \label{eq:amu_gen}
\Delta a_\mu&=&\dfrac{\chi_{a_\mu} m_\mu^2}{8 \pi^2 m_{\Phi_M}^2}\,
\left[
\left( L^{\mu *}_{AM} L^\mu_{AM}  + R^{\mu *}_{AM} R^\mu_{AM}\right)
\left(Q_{\Phi_M} \widetilde F_7\left(x_{AM}\right) - Q_{\Psi_A} F_7\left(x_{AM}\right)\right)\right. \\
&&\qquad\qquad\quad +
\left. 
\left( L^{\mu *}_{AM} R^\mu_{AM}+R^{\mu *}_{AM} L^\mu_{AM} \right)
\dfrac{2 m_{\Psi_A}}{m_\mu}
\left(Q_{\Phi_M} \widetilde G_7\left(x_{AM}\right) - Q_{\Psi_A} G_7\left(x_{AM}\right)\right)
\right]\,, \nonumber\\
d_{\mu}&=&\dfrac{\chi_{a_\mu} m_{\Psi_A}}{8 \pi^2 m_{\Phi_M}^2}\,
e \left( L^{\mu *}_{AM} R^\mu_{AM}-R^{\mu *}_{AM} L^\mu_{AM} \right)
\left(Q_{\Phi_M} \widetilde G_7\left(x_{AM}\right) - Q_{\Psi_A} G_7\left(x_{AM}\right)\right)\,, 
\label{eq:amu5_gen}
\eea 
where $m_\mu$ is the muon mass, $\chi_{a_\mu}$ is the colour factor given in Table~\ref{tab:SU3_gm2}, and $Q_{\Phi_M}$ and $Q_{\Psi_A}$ are the electric charges of the NP fields ${\Phi_M}$ and ${\Psi_A}$, respectively. Analogously to photon-penguin contributions to $b \to s$ transitions, the conservation of electric charge imposes that $Q_{\Phi_M} + Q_{\Psi_A} = Q_\mu \equiv -1$. Finally, the loop functions $F_7(x)$, $\widetilde F_7(x)$, $G_7(x)$ and $\widetilde G_7(x)$ are defined in Appendix~\ref{app:loopfunctions}.


\boldmath
\subsection{Modified \texorpdfstring{$Z$}{Z} Couplings}
\label{sec:Zmumu}
\unboldmath

\begin{figure}[!t]
\centering
\includegraphics[scale=0.45]{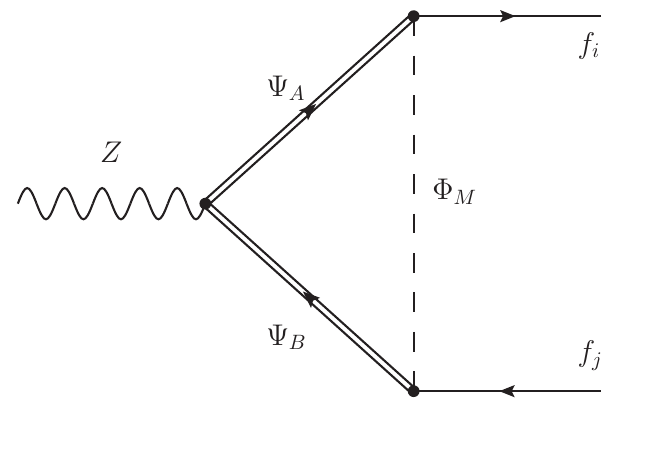}
\includegraphics[scale=0.45]{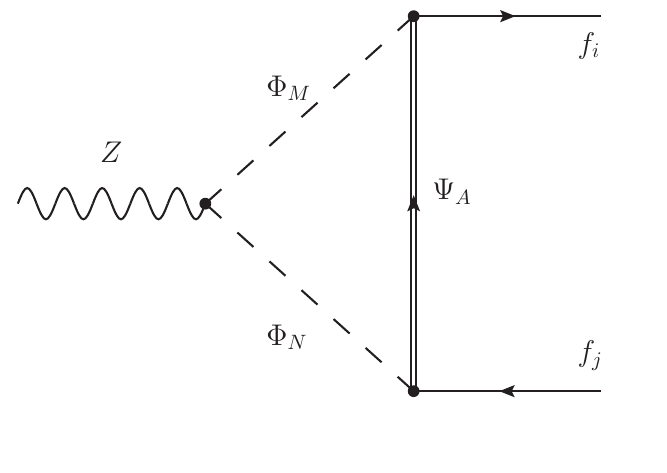}
\includegraphics[scale=0.45]{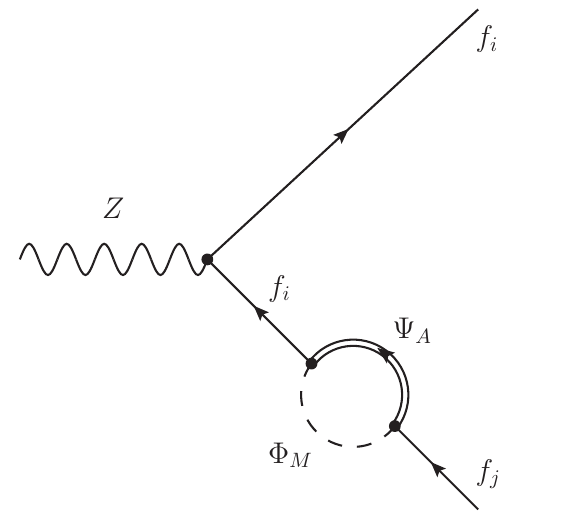}
\caption{Feynman diagrams modifying the $Z \bar{f_i} f_j$ vertex with $f_i=s,\,b,\,\mu$.}
\label{fig:diagZff}
\end{figure}

Here, we study the effects of our new particles on modified $Z$ couplings, i.e. on $Z\bar\mu\mu$, $Z\bar bb$, $Z\bar ss$ and $Z\bar s b$ couplings, both for off- and on-shell $Z$ bosons\footnote{Expressions for $Z$ couplings in generic gauge theories can be found in Ref.~\cite{Brod:2019bro}.}. We define the form-factors governing $Z\bar ff$ interactions as~\cite{Dedes:2017zog}
\beq\label{eq:leff-z}
-\dfrac{g_2}{c_W}\bar{f^\prime} \gamma^\mu
\Big{[}g^{f^\prime f}_L(q^2)\,P_L+ g^{f^\prime f}_R(q^2) \,P_R \Big{]} f\, Z_\mu+ \text{h.c.} \,,
\eeq
where $f=\{b,s,\,\mu\}$, $g_2$ is the $SU(2)$ gauge coupling, $\theta_W$ the Weinberg angle and $q$ is the $Z$ momentum. Moreover, 
\begin{equation}
g^{f^\prime f}_{L(R)}(q^2)= g_{f_L}^{\mathrm{SM}} \delta _{f^\prime f} +\Delta g^{f^\prime f}_{L(R)}(q^2)
\end{equation}
with  $g_{f_L}^{\mathrm{SM}}=(T_3^f-Q_f s^2_W)$ and $g_{f_R}^{\mathrm{SM}}=-Q_f s^2_W$ being the $Z$ couplings to SM fermions at  tree-level.
The relevant Feynman diagrams are shown in Fig.~\ref{fig:diagZff}. We write the coupling of the $Z$ boson to the new scalars and fermions as
\beq\label{eq:LZff}
\mathcal{L}^Z = -\dfrac{g_2}{c_W}Z_\mu \left(\bar{\Psi}_A \gamma^\mu
\Big{[}g^{\Psi, L}_{AB}\,P_L+ g^{\Psi, R}_{AB} \,P_R \Big{]} \Psi_B  +  g^{\Phi}_{MN} \,
\Phi_M^\dagger \,i\overset{\leftrightarrow}{\partial^\mu} \,\Phi_N \, \right) + \hc
\,,
\eeq
where we have introduced the notation $a\overset{\leftrightarrow}{\partial^\mu}b = a({\partial^\mu}b) - ({\partial^\mu} a)b$, and with generic couplings $g^{\Psi, L,R}_{AB}$ and $g^{\Phi}_{MN}$ which can only be determined in a UV complete model in which also the couplings of the new particles to the SM Higgs are known.
Using the generic Lagrangian from Eq.~(\ref{eq:L_bsmumu}), one obtains the following results for the coefficients
\bea
\Delta g^{f^\prime f}_{L}(q^2)  &=&
 \dfrac{ \chi_Z \,L^{f^\prime}_{BN} L^{f*}_{AM} }{{32\pi^2}} \nonumber\\
&& \qquad\left[ 
2\,g_{AB}^{\Psi, L} \delta_{MN} \dfrac{m_{\Psi_A} m_{\Psi_B}}{m_{\Phi_M}^2} 
G_Z(x_{AM},x_{BM}) 
-g_{AB}^{\Psi, R} \delta_{MN} F_Z(x_{AM},x_{BM},m_{\Phi_M}) 
 \right. \nonumber\\
&& \qquad +
g_{MN}^{\Phi} \delta_{AB} H_Z(x_{AM},x_{AN},m_{\Psi_A})
-\dfrac{1}{2}(g_{f_L}^{SM} +g_{f^\prime_L}^{SM}) \delta_{AB} \delta_{MN}  I_Z(x_{AM},m_{\Phi_M}) 
\nonumber\\
&& \qquad+\left. 
q^2 \left( 
g_{AB}^{\Psi, L} \delta_{MN} \dfrac{m_{\Psi_A} m_{\Psi_B}}{m_{\Phi_M}^4}  \widetilde G_Z(x_{AM},x_{BM}) 
-\dfrac{2}{3} \dfrac{g_{AB}^{\Psi, R} \delta_{MN} }{m_{\Phi_M}^2} \widetilde F_Z(x_{AM},x_{BM})\right.\right. \nonumber\\
&& \qquad \left.\left.\hspace*{1.5cm}
-\dfrac{1}{3} \dfrac{g_{MN}^{\Phi} \delta_{AB}}{m^2_{\Psi_A}} \widetilde  H_Z(x_{AM},x_{AN})\right.\Big)  \right.\Big]\label{eq:gLZff_gen}\,, \\
\Delta g^{f^\prime f}_{R}(q^2)  &=& \Delta g^{f^\prime f}_{L}(q^2)\left(L\leftrightarrow R \right)
\,, \label{eq:gRZff_gen}
\eea 
where the loop functions are defined in Appendix~\ref{app:loopfunctions}, and the colour factor $\chi_Z=\chi_\gamma$ for $f,f^\prime=b,s$ (see Table~\ref{tab:SU3_bsg}) and $\chi_Z=\chi_{a_\mu}$ (see Table~\ref{tab:SU3_gm2}) for $f=\mu$. Here we have set the masses and momenta of the external fermions to 0 and expanded up to first order in $q^2$ over the NP scale. If one is considering data from $Z$ decays, \eq{eq:gLZff_gen} has to be evaluated to $q^2=m_Z^2$ while for processes with an off-shell $Z$ (like $b\to s\ell^+\ell^-$) one has to set $q^2=0$.
\medskip

Note that in the absence of EW symmetry breaking in the NP sector, the contribution of the self-energies cancel the one of the genuine vertex correction and \eq{eq:gLZff_gen} vanishes for $q^2=0$. Therefore, as noted above, $g^{\Psi, L}_{AB}$, $g^{\Psi, R}_{AB}$ and $g^{\Phi}_{MN}$ are only meaningful after EWSB and it is not possible to relate them purely to $SU(2)\times U(1)$ quantum numbers. In a specific UV model with a known pattern of EWSB, rotation matrices can be used to relate the couplings before and after the breaking. Consequently, the cancellation of UV divergences (present in some of the loop functions in \eq{eq:gLZff_gen}) is only manifest after summation over $SU(2)$ indices, due to a GIM-like cancellation originating from the unitarity of the rotation matrices. We will give a concrete example of this in Sec.~\ref{sec:4thgen}.
\medskip

The form-factors in Eq.~\eqref{eq:leff-z} includes $Z\bar sb$ couplings generating contributions to $C_{9,{10}}^{(\prime)}$
\bea
C_9^{(\prime)\,Z}&=& \mathcal{N}\dfrac{ \pi g^2 }{\alpha_{\rm EM}c^2_{W} m_Z^2} g^{sb}_{{L(R)}}(q^2=0) \left(1 - 4 s_W^2 \right)\,, \\
C_{10}^{(\prime)\,Z}&=& -\mathcal{N}\dfrac{\pi g^2 }{\alpha_{\rm EM} c^2_{W} m_Z^2}  g^{sb}_{{L(R)}}(q^2=0) \,.
\eea 
Note that these contributions are lepton flavour universal and therefore cannot account for $R_K$ and $R_{K^*}$. However, a mixture of lepton flavour universal and violating contributions is phenomenologically interesting~\cite{Alguero:2018nvb}, especially in the light of the recant Belle and LHCb measurements~\cite{Alguero:2019ptt,Datta:2019zca}.
In a similar fashion, $Z\bar sb$ couplings will also generate the following contributions to $b\to s\nu\bar\nu$ contained in $C_{L(R)}$
\bea
C_{L(R)}^{\,Z}&=& -\mathcal{N}\dfrac{ \pi g^2 }{\alpha_{\rm EM}c^2_{W} m_Z^2 } g^{sb}_{L{(R)}}(q^2=0)\,.
\eea 
\medskip

Finally,  if $SU(2)$ invariance at the NP scale is imposed, the new scalars and fermions couple also the neutrinos.
Hence, contributions to $Z\to \nu\bar\nu$ and $W\to \mu\bar \nu$ will arise as well.  Concerning $Z\to \nu\bar\nu$, $g_{\nu_L}(q^2=m_Z^2)$ can be straightforwardly extracted from Eq.~(\ref{eq:gLZff_gen}) by appropriate replacements 
and the same is true concerning  $W\mu \bar\nu$ couplings.
\medskip

\section{Experimental Constraints on Wilson Coefficients}\label{sec:Exp}

In this section we review the experimental situation and the resulting constraints on the Wilson coefficients calculated in the previous section.
\medskip

\boldmath
\subsection{\texorpdfstring{$b\to s$}{btos} Transitions}\label{sec:Expbsmumu}
\unboldmath

The semileptonic operators $\mathcal{O}_{9,10}^{(\prime)}$, $\mathcal{O}_{S,P}^{(\prime)}$ and $\mathcal{O}_{T}^{(\prime)}$, together with magnetic operators $\mathcal{O}_{7}^{(\prime)}$, contribute to a plethora of $b\to s\ell^+\ell^-$ observables. The corresponding measurements include total branching ratios of $B_s\to \ell^+\ell^-$~\cite{Amhis:2016xyh}, of the exclusive decays $B \to K^* \gamma$~\cite{Amhis:2016xyh}, $B \to \phi \gamma$~\cite{Aaij:2012ita}, the inclusive decay $B\to X_s\gamma$~\cite{Amhis:2016xyh}, the angular analyses of  $B\to K^{(*)}\ell^+\ell^-$~\cite{Aaij:2015oid,Aaij:2015dea,Aaij:2016flj,Wehle:2016yoi,Aaboud:2018krd,Khachatryan:2015isa,Sirunyan:2017dhj} (proposed in Refs.~\cite{Altmannshofer:2008dz,Matias:2012xw,Descotes-Genon:2013vna}) and $B_s \to \phi \, \ell^+ \ell^-$~\cite{Aaij:2015esa}, and also the ratios $R_K$~\cite{Aaij:2019wad} and $R_{K^*}$~\cite{Aaij:2017vbb,Abdesselam:2019wac} measuring lepton flavour universality violation.
\medskip

First of all, the contributions of scalar operators are helicity-enhanced in the $B_s\to\mu^+\mu^-$ branching ratio with respect to the $O_{10}$ contribution of the SM. This results in the bound~\cite{Beneke:2017vpq}
\bea
\label{eq:BS}
\dfrac{{{\mathcal B}^{\rm exp}}(B_s\to \mu^+\mu^-)}{{\mathcal B^{\rm SM}}(B_s\to \mu^+\mu^-)} -1&=& 
\left|1+\dfrac{C_{10}-C_{10}^\prime}{C_{10}^{\text{SM}}} 
+ \dfrac{m_{B_s}^2}{2 m_\mu(m_b+m_s)}\dfrac{C_P-C_{P}^\prime}{C_{10}^{\text{SM}}} \right|^2 \nonumber \\
&& +\dfrac{m_{B_s}^2(m_{B_s}^2-4 m_\mu^2)}{4 m_\mu^2(m_b+m_s)^2}\left|\dfrac{C_S-C_S^\prime}{C_{10}^{\text{SM}}} \right|^2 - 1 = -0.13 \pm 0.20\,,\quad
\eea
which excludes sizable contributions to scalar operators (unless there is a purely scalar quark current) and leads to
\bea
|C_{S,P}^{(\prime)} |\lesssim  0.03 \,\quad (2\,\sigma)\,,
\eea
from the updated one-parameter fit of Ref.~\cite{Altmannshofer:2017wqy}. Therefore, we neglect the effects of scalar operators in semi-leptonic $B$ since they anyway cannot explain the corresponding anomalies.
\medskip

Moreover, the inclusive $b\to s\gamma$ decay strongly constrains the magnetic operators. From~\cite{Misiak:2017woa, Misiak:2015xwa}, in the limit of vanishing $C_{7,8}^\prime$\footnote{Note that $C_{7,8}^\prime$ are less constrained since they do not interfere with the SM. For a more detailed analysis including primed operators see e.g. 	Ref.~\cite{Hurth:2003dk}.}, we have 
\bea
\dfrac{{{\mathcal B}^{\rm exp}}(b\to s\gamma)}{{\mathcal B^{\rm SM}}(b\to s\gamma)}-1=- 2.87\, \left[C_7+0.19\,C_8\right]=(-0.7\pm 8.2)\times 10^{-2}\,,
\eea
leading to
\bea
|C_7+0.19\,C_8 | \lesssim 0.06\,\quad (2\,\sigma)\,.
\eea
Here, we used $C_{7,8}$ at a matching scale of 1 TeV as input. Again, these constraints are so stringent that the effect of $C_{7,8}$ on the flavour anomalies can be mostly neglected.
\medskip 

On the other hand, vector operators can explain the $b\to s \ell^+\ell^-$ anomalies. We therefore refer to global fits to constrain $C_{9,10}^{(\prime)}$, where all the relevant observables have been taken into account~\cite{Capdevila:2017bsm,Altmannshofer:2017yso,DAmico:2017mtc,Hiller:2017bzc,Geng:2017svp,Ciuchini:2017mik,Alok:2017sui,Hurth:2017hxg}. The results of the most recent fits find at the $2\,\sigma$ level 
\bea
\begin{aligned}
-1.48 &\leq  C_{9} \leq -0.71\,, \qquad
-0.12 \leq C_{10} \leq 0.61\,,  \nonumber\\
-0.56 &\leq  C_{9}^{\prime} \leq 1.14\,, \qquad\quad
-0.57 \leq C_{10}^{\prime} \leq 0.34\,, \qquad
\end{aligned}
\eea
according to Ref.~\cite{Alguero:2019ptt} (which is compatible with Refs.~\cite{Alok:2019ufo,Ciuchini:2019usw,Aebischer:2019mlg,Kowalska:2019ley}).
\medskip

As explained in Sec.~\ref{sec:BKnunu}, $SU(2)$ invariance implies the presence of contributions to $B \to K^{(*)} \nu\bar\nu$ decays as well. Since there is no experimental way to distinguish different neutrino flavours in these decays, one measures the total branching ratio which we normalize to its SM prediction~\cite{Buras:2014fpa}:
\beq
R_{K^{(*)}}^{\nu\bar\nu} = \frac{{\mathcal B}^{\rm exp}(B\to K^{(*)} \nu\bar\nu)}{{\mathcal B}^{\rm SM}(B\to K^{(*)} \nu\bar\nu)} =  
\frac{2(C_L^{\rm SM})^2 + (C_L^{\rm SM} + C_L)^2 - \kappa\ (C_L^{\rm SM} + C_L)C_R + C_R^2}{3(C_L^{\rm SM})^2}  \,.
\eeq
In the case of a $K$ in the final state one has $\kappa \equiv -2$, while for the $K^*$ one gets $\kappa = 1.34(4)$~\cite{Buras:2014fpa}. The current experimental limits at $90 \%$ C.L. are~\cite{Grygier:2017tzo}
\beq
R_{K}^{\nu \bar \nu} < 3.9 \,, \qquad\qquad R_{K^*}^{\nu \bar \nu} < 2.7 \,.
\eeq

\boldmath
\subsection{Neutral Meson Mixing}
\unboldmath

The experimental constraint on the Wilson coefficients in Eqs.~\eqref{eq:CBB_gen_beg}-\eqref{eq:CBB_gen_end} comes from the mass difference $\Delta M_s$ of neutral $B_s$ mesons, see e.g. Ref.~\cite{Gabbiani:1996hi} (in the case of real Wilson coefficients). To compare our results with experiments we use
\bea\label{eq:dmsnum}
\! \!\!\!\frac{\Delta M_s^{\rm exp}}{\Delta M_s^{\rm SM}}  &=&
\left\vert 1 +
\sum_{i,j=1}^3  R_i(\mu_b)
\frac{\eta_{ij}(\mu_b,\mu_H)}{C^{\rm SM}_1(\mu_b)}\left(C_j+\widetilde C_j\right)
+ \sum_{i,j=4}^5  R_i(\mu_b)
\frac{\eta_{ij}(\mu_b,\mu_H)}{C^{\rm SM}_1(\mu_b)} C_j\right\vert \nonumber\\
&=&
\left\vert 1 +  \dfrac{0.8\,(C_1  + \widetilde C_1)
 -1.9\,  (C_2  + \widetilde C_2) + 0.5\,(C_3  + \widetilde C_3)  + 
 5.2\,C_4  +  1.9\, C_5}{C^{\rm SM}_1(\mu_b)}
\right\vert \,,\
\eea
where $R_i(\mu_b)$ is related to the matrix element of the operators $Q_i$ in Eq.~\eqref{eq:OPBB} at the scale $\mu_b$ by the relation 
\beq
R_i(\mu_b)=\frac{\langle \bar{B}_s| Q_i(\mu_b)| B_s \rangle}{\langle \bar{B}_s| Q_1(\mu_b)| B_s \rangle}\,.\label{eq:matrix_elements}
\eeq
The coefficients $C_i$ and  $\widetilde{C}_i$ are the ones in Eqs.~\eqref{eq:CBB_gen_beg}-\eqref{eq:CBB_gen_end}, computed at the NP scale $\mu_H$. The matrix in operator space $\eta_{ij}(\mu_b,\mu_H)$  encodes the QCD evolution from the high scale $\mu_H$ to $\mu_b$, which we calculated numerically for a reference scale $\mu_H=1 $ TeV~\cite{Becirevic:2001jj}. The matrix elements in Eqs.~\eqref{eq:dmsnum}-\eqref{eq:matrix_elements} have been computed by a $N_f=2+1$ lattice simulation~\cite{Bazavov:2016nty}, which found values consistent with the $N_f=2$ calculation~\cite{Carrasco:2013zta} and recent sum rules results~\cite{King:2019lal}. It is worth mentioning that FLAG-2019~\cite{Aoki:2019cca} only provides a lattice average for $\langle \bar{B}_s| Q_1(\mu_b)| B_s \rangle$, which is however dominated by the $N_f=2+1$ results from Ref.~\cite{Bazavov:2016nty}. Therefore, we decided to employ the results from Ref.~\cite{Bazavov:2016nty} in Eqs.~\eqref{eq:dmsnum}-\eqref{eq:matrix_elements}. The SM value for the Wilson coefficient is $C^{\rm SM}_1(\mu_b)=\dfrac{G_F^2 M_W^2}{4 \pi ^2}\lambda_t^2  \eta_{11}(\mu_b,m_t) S_0(x_t)\simeq 7.2\times10^{-11} \text{ GeV}^{-2} $.
\medskip

The experimental constraint therefore reads~\cite{Tanabashi:2018oca} 
\beq
R_{\Delta M_s} = \frac{\Delta M_s^{\rm exp}}{\Delta M_s^{\rm SM}}  - 1 =-0.09\pm 0.08\,,
\eeq 
computed with the values from Ref.~\cite{Bazavov:2016nty} for $\langle \bar{B}_s| Q_1(\mu_b)| B_s \rangle$. This value shows a slight tension with the SM as first outlined in Refs.~\cite{DiLuzio:2018wch,DiLuzio:2017fdq}. The tension would be reduced if one considered the results for the matrix element from Ref.~\cite{King:2019lal}: however, in this case one should rely on a separate computation for the decay constant, while in Ref.~\cite{Bazavov:2016nty} both quantities are computed together.
\medskip

Analogously to the $B_s$ system, $D_0 - \bar{D}_0$ mixing is constrained by the mass difference of neutral $D_0$ mesons~\cite{Tanabashi:2018oca}:
\beq
\Delta M_{D_0}^{\rm exp} = (0.63^{+0.27}_{-0.29}) \times 10^{-11} {\rm MeV}\,.
\eeq
Unfortunately, a precise SM prediction is still lacking in this sector but one can constrain the NP contribution by assuming that not more than the total mass difference is generated by it.
\medskip

\subsection{Anomalous Magnetic Moment of the Muon}

From the experimental side, this quantity has been already measured quite precisely~\cite{Bennett:2006fi}, but further improvements by experiments at Fermilab~\cite{Grange:2015fou} and J-PARC~\cite{Saito:2012zz} (see also~\cite{Gorringe:2015cma}) are expected in the future. On the theory side, the SM prediction has been improved continuously~\cite{Czarnecki:1995wq,Czarnecki:1995sz,Jegerlehner:2009ry,Hagiwara:2011af,Aoyama:2012wk,Gnendiger:2013pva,Chakraborty:2016mwy,Jegerlehner:2017lbd,DellaMorte:2017dyu,Davier:2017zfy,Borsanyi:2017zdw,Giusti:2018mdh,Giusti:2019xct,Blum:2018mom,Keshavarzi:2018mgv,Colangelo:2015ama,Green:2015sra,Gerardin:2016cqj,Blum:2016lnc,Colangelo:2017qdm,Blum:2017cer,Hoferichter:2018dmo,Kurz:2014wya,Colangelo:2014qya}. The current tension between the two determinations accounts to
\beq
\Delta a_\mu=a_\mu^{exp} - a_\mu^\text{SM}\sim 270(85)\times 10^{-11}\,.
\eeq

\begin{boldmath}
\subsection{\texorpdfstring{$Z$}{Z} Decays}
\end{boldmath}

The main experimental measurements of $Z$ couplings have been performed at LEP~\cite{ALEPH:2005ab} (at the $Z$ pole). 
We extract from the model independent analysis of Ref.~\cite{Efrati:2015eaa} the values for the NP 
contributions~\footnote{$Z$ couplings in Ref.~\cite{Efrati:2015eaa} are defined with opposite sign with respect to our conventions~\cite{Dedes:2017zog}.}
\beq
\begin{aligned}
\Delta g_{\mu_{L}}(m_Z^2) &= -(0.1 \pm 1.1)\times 10^{-3} \,, \qquad\qquad 
&\Delta g_{\mu_{R}}(m_Z^2) =&\ (0.0 \pm 1.3)\times 10^{-3} \,, \\
\Delta g_{b_{L}}(m_Z^2)  &= -(0.33 \pm 0.16)\times 10^{-2} \,, 
&\Delta g_{b_{R}}(m_Z^2)  =&\ -(2.30 \pm 0.82)\times 10^{-3} \,, \\
 \Delta g_{\nu_{L}}(m_Z^2)  &=(0.40 \pm 0.21)\times 10^{-2}\,, &&
\end{aligned}
\eeq
neglecting cancellations and correlations.


\begin{boldmath}
\section{\texorpdfstring{${4^{\rm th}}$}{4th} Generation Model}\label{sec:4thgen}
\end{boldmath}

In this section we propose a model with a vector-like 4${}^{\rm th}$ generation of fermions and a new complex scalar. This will also allow us to apply and illustrate the generic findings of the previous section to a UV complete model and study the effects in $b\to s\ell^+\ell^-$ data and $a_\mu$.
\medskip

\subsection{Lagrangian}
The Lagrangian for our  4${}^{\rm th}$ generation model is obtained from the SM one by adding a 4${}^{\rm th}$ vector-like generation~\cite{Poh:2017tfo,Raby:2017igl}  and a neutral scalar
\begin{eqnarray}\label{eq:L_4th}
L^{4{\rm th }} &=& \sum\limits_{i } \left( \Gamma^L_{q_i} \bar \Psi_q P_L q_i + \Gamma^L_{\ell_i} \bar \Psi_\ell P_L \ell_i  + \Gamma^R_{u_i} \bar \Psi_u P_R u_i + \Gamma^R_{d_i} \bar \Psi_d P_R d_i + \Gamma^R_{e_i} \bar \Psi_e P_R e_i  \right) \Phi + \hc \nonumber\\
&& + \sum\limits_{C = L,R} \left( \lambda_C^U \bar\Psi_q P_C \tilde h \Psi_u + \lambda_C^D \bar\Psi_q P_C h \Psi_d + \lambda_C^E \bar\Psi_\ell P_C h \Psi_e  \right) + \hc \nonumber\\
&& + \sum\limits_{F = q,\ell,u,d,e} M_F \bar\Psi_F \Psi_F+ \kappa\, h^\dag h\, \Phi^\dag\Phi + m_\Phi^2 \Phi^\dag\Phi\,,
\end{eqnarray}
where $i$ is a family index and $h$ the SM Higgs doublet. The charge assignments for the new vector-like fermions $\Psi=\Psi_L+\Psi_R$ with $P_{L,R}\Psi=\Psi_{L,R}$ and the new scalar $\Phi$ are
\begin{equation}
\begin{array}{*{20}{c}}
{}&\vline& {SU\!\left(3 \right)}&{SU\!\left( 2 \right)}&{U{{\left( 1 \right)}}} & U'\left( 1 \right)\\
\hline
\Psi_q&\vline& 3&2&{1/6} & Z\\
\Psi_u&\vline& 3&1&{2/3} & Z\\
\Psi_d&\vline& 3&1&{- 1/3} & Z\\
\Psi_\ell&\vline& 1&2&{- 1/2} & Z\\
\Psi_e&\vline& 1&1&{- 1} & Z\\
\Phi&\vline& 1&1& 0 & -Z
\end{array}\,.
\label{table4th}
\end{equation}
The SM fermions have the same $SU(3)\times SU(2) \times U(1)$ charge assignments of the relative NP fermion partner, and the higgs transforms as a (1, 2, 1/2). Here we assigned to the new particles also charges under a new $U(1)$ group in order to forbid mixing with the SM particles, giving a similar effect as R-parity in the MSSM\footnote{We did not assume a $Z_2$ symmetry because this would allow the scalar $\Phi$ to be real and lead to crossed boxes in $b\to s\ell^+\ell^-$, canceling the desired effect there.}. From Table~\ref{table4th} we see that concerning $SU(3)$ we are dealing with cases I of our generic analysis in Tables~\ref{tab:SU3_bsmumu}-\ref{tab:SU3_gm2}. In particular, concerning $b \to s \mu^+ \mu^-$, this model would generate diagrams of type a) in Fig.~\ref{fig:diagbsmumu}.
\medskip

After EWSB, mass matrices for the new fermions are generated
\bea
L^{4{\rm th }}_{\text{mass}} &=& {\left( {\begin{array}{*{20}{c}}
		\bar \Psi_{q,1}\\
		{\bar \Psi}_u
		\end{array}} \right)^T}{{\bf{M}}_U}{P_L}\left( {\begin{array}{*{20}{c}}
	\Psi_{q,1}\\
	{{\Psi_u}}
	\end{array}} \right) 
	+ {\left( {\begin{array}{*{20}{c}}
		\bar \Psi_{q,2}\\
		{\bar \Psi }_d
		\end{array}} \right)^T}{{\bf{M}}_D}{P_L}\left( {\begin{array}{*{20}{c}}
	\Psi_{q,2}\\
	{{\Psi_d}}
	\end{array}} \right) \nonumber\\
&&	+ {\left( {\begin{array}{*{20}{c}}
		\bar \Psi_{\ell,2}\\
		{{{\bar \Psi }_e}}
		\end{array}} \right)^T}{{\bf{M}}_E}{P_L}\left( {\begin{array}{*{20}{c}}
	\Psi_{\ell,2}\\
	{{\Psi_e}}
	\end{array}} \right) + \hc\,,
\eea
where  ${{\bf{M}}_{U,D,E}}$ are non-diagonal mass matrices  
\beq\label{eq:MUDE}
{{\bf{M}}_{D(U)}} = \left( {\begin{array}{*{20}{c}}
		{{M_q}}&{\sqrt 2 v\lambda _R^{D(U)}}\\
		{\sqrt 2 v\lambda _L^{{D(U)}*}}&{{M_{d(u)}}}
\end{array}} \right)\,, \quad
{{\bf{M}}_E} = \left( {\begin{array}{*{20}{c}}
		{{M_\ell}}&{\sqrt 2 v\lambda _R^E}\\
		{\sqrt 2 v\lambda _L^{E*}}&{{M_e}}
\end{array}} \right)\,.
\eeq
Here the subscripts 1 and 2 denote the $SU(2)$ component of the doublet. We diagonalize these mass matrices by performing the field redefinitions
\bea\label{eq:field_rot}
P_L {\left( {\begin{array}{*{20}{c}}
		\Psi_{q,1}\\
		{\Psi _u}
		\end{array}} \right)_I} \to W^{U_L}_{IJ}\Psi^{{U_L}}_J\,,\qquad
P_L {\left( {\begin{array}{*{20}{c}}
		\Psi_{q,2}\\
		{\Psi _d}
		\end{array}} \right)_I} \to W^{D_L}_{IJ}\Psi^{{D_L}}_J\,,\qquad
+ \quad  L \to R \nonumber\\
P_L \Psi_{L,1} \to \Psi ^{{N_L}} \,,\qquad
P_L {\left( {\begin{array}{*{20}{c}}
		\Psi_{\ell,2}\\
		{\Psi _e}
		\end{array}} \right)_I} \to W^{E_L}_{IJ}\Psi^{{E_L}}_J\,,\qquad
+ \quad  L \to R 
\eea
leading to
\beq \label{eq:M_rotation}
(W^{{F_L}\dagger}{\bf{M}}_FW^{{F_R}})_{IJ} = m_{F_I}\ {\delta _{IJ}}\,,\qquad{\rm with}\;F=U,D,E\,.
\eeq
Therefore, after EWSB we have the mass eigenstates $\Psi^{U_{L,R}}_I$, $\Psi^{D_{L,R}}_I$, $\Psi^{E_{L,R}}_I$ and $\Psi^{N_{L,R}}$, with $I=\{1,2\}$. In particular, $\Psi^{U_{L,R}}_I$ and $\Psi^{D_{L,R}}_I$ ($\Psi^{E_{L,R}}_I$ and $\Psi^{N_{L,R}}$) are $SU(3)$ triplets (singlets) with the same electric charges as up-type and down-type quarks (charged-leptons and neutrinos), respectively.
\medskip

The rotations introduced at Eq.~\eqref{eq:field_rot} lead to the following Lagrangian for the interactions in the broken phase
\bea\label{eq:LR4thint}
L_{\rm int}^{{\text{4th}}} &=& \left(L_I^{d_i} \bar \Psi _I^{D} P_L d_i  
+ L_I^{e_i} \bar \Psi _I^{E} P_L e_i 
+ R_I^{d_i} \bar \Psi _I^{D} P_R d_i  
+ R_I^{e_i} \bar \Psi _I^{E} P_R e_i \right)  \Phi \nonumber\\
&&+ \left( L_I^{u_i} \bar \Psi _I^{U} P_L u_i 
+ L^{\nu_i} \bar \Psi^{N} P_L \nu_i   
+ R_I^{u_i} \bar \Psi _I^{U} P_R u_i \right)  \Phi  + \hc\,
\eea
which resembles \eq{eq:L_bsmumu} for the special case of our 4th generation model. Thus identify 
\bea\label{eq:LRGamma}
L_I^{d_i}  &=&\Gamma^L_{q_i} W_{1I}^{{D_R}^*}\,,\qquad 
L_I^{e_i} = \Gamma^L_{\ell_i} W_{1I}^{{E_R}^*}\,,\qquad 
L_I^{u_i}  =\Gamma^L_{q_j} V_{ij}^{*}W_{1I}^{{U_R}^*} \,,\qquad 
L^{\nu_i} = \Gamma^L_{\ell_i} \,, \nonumber\\
R_I^{d_i}  &=&\Gamma^R_{d_i} W_{2I}^{{D_L}^*}\,,\qquad 
R_I^{e_i} = \Gamma^R_{e_i} W_{2I}^{{E_L}^*}\,,\qquad
R_I^{u_i} =\Gamma^R_{u_i} W_{2I}^{{U_L}^*}\,.
\eea
Here we worked in the down-basis for the SM quarks which means that CKM matrices $V_{ij}$ appear in vertices involving up-type quarks. The first two columns of the above Lagrangian involves couplings with down-type quarks and charged leptons and can be directly matched on the Lagrangian in Eq.~(\ref{eq:L_bsmumu}) for the case of only one scalar, i.e. $\Phi_M \equiv \Phi$ and  $\Psi_A \equiv \{\Psi_I^D, \Psi_I^E\}$. The presence of $L_I^{u_i}$ ($L^{\nu_i}$) resembles the fact, mentioned in Sec.~\ref{sec:generic}, that left-handed couplings to down-quarks (leptons) lead via $SU(2)$ to couplings to left-handed up-quarks (neutrinos). In addition couplings to right-handed up-quarks $R_I^{u_i}$ appear in our model which are however not relevant for our phenomenology.
\medskip


\begin{boldmath}
\subsection{Wilson Coefficients}
\end{boldmath}
\label{sec:numerics}
With these conventions we can now easily derive the Wilson coefficients within our model which can be directly obtained from the results of Sec.~\ref{sec:generic}. In order to simplify the expressions, we will assume $M_Q=M_d \equiv m_D$ and $M_L = M_e \equiv m_E$ and only take into account couplings to $b$, $s$ and $\mu$ in Eq.~\eqref{eq:L_4th}:
\bea
\{\Gamma_s^L\,, \Gamma_b^L \,, \Gamma_\mu^L \,, \Gamma_s^R \,, \Gamma_b^R \,, \Gamma_\mu^R \}\,,
\eea
Concerning $SU(2)$ breaking effects the couplings $\lambda_{L,R}^{D}$ and $\lambda_{L,R}^{E}$ related to the down and charged leptons sector, respectively, can be relevant. However, concerning $\lambda_{L,R}^{D}$ recall that from Section~\ref{sec:Expbsmumu} that experimental data suggests very small values for $C_{S,P}$ and $C_{7,8}$. In our model this can be achieved by assuming $\lambda_{L,R}^{D} = 0$\footnote{Note that the effect in scalar and magnetic operators can also be suppressed if $\Gamma^{R}_{b,s}=0$ or very small. However, we decided to focus on option with $\lambda_{L,R}^{D}$ being very small.}. In this limit the mass matrix ${\bf M}^D$ in Eq.~\eqref{eq:MUDE} is diagonal and the corresponding rotation matrices $W^{{D_{R(L)}}}$ in Eq.~\eqref{eq:M_rotation} are equal to the identity, which implies
\beq
C_{S,P} \propto L_A^{s*}R_A^b \propto W_{1A}^{{D_R}} W_{2A}^{{D_L}^*} = \delta_{1A}\delta_{2A} = 0\,.
\eeq
With this setup, we obtain the following non-vanishing couplings in the quark sector of the Lagrangian in Eq.~\eqref{eq:LR4thint}:
\bea\label{eq:LR_4th}
L_1^s &=& \Gamma_s^L \,, \qquad 
L_1^b = \Gamma_b^L \,,\qquad
R_2^s = \Gamma_s^R \,, \qquad 
R_2^b = \Gamma_b^R \,, \nonumber\\
L_1^u &=& V_{us}^*\ \Gamma_s^L + V_{ub}^*\ \Gamma_b^L \,, \qquad
L_1^c = V_{cs}^*\ \Gamma_s^L + V_{cb}^*\ \Gamma_b^L \,.
\eea
with
\begin{eqnarray}
\Gamma^L\equiv L^b_1 L_1^{s*} \,, \qquad 
\Gamma^R\equiv R^b_2 R_2^{s*} \,, \qquad 
x_{D(E)}\equiv \dfrac{m_{D(E)}^2}{m_\Phi^2}\,.
\end{eqnarray}

The expressions of Wilson coefficients for $b\to s$ processes simplify to:
\medskip 

\noindent $\bullet$ $b \to s \mu^+\mu^-$ and $b \to s \gamma$ (see Eqs.~(\ref{eq:C9_gen})-(\ref{eq:C8g_gen}))
\bea
C_9^{{\text{box}}\,} &=& 
-{\cal N}\frac{\Gamma^L }{{32\pi \alpha_{\text{EM}} m_{\Phi}^2}}  
\left( |\Gamma^L_\mu|^2 + |\Gamma^R_\mu|^2 \right) F(x_D, x_E)   \,, \\
C_{10}^{{\text{box}}\,} &=&
{\cal N}\frac{\Gamma^L }{{32\pi \alpha_{\text{EM}} m_{\Phi}^2}}
\left( |\Gamma^L_\mu|^2 - |\Gamma^R_\mu|^2 \right)F(x_D, x_E) \,, \\
C_9^{\gamma} &=& {\cal N} \dfrac{ \Gamma^L}{6m_\Phi^2} \widetilde G_9\left(x_D\right)\,,\:\:
C_7= {\cal N} \dfrac{ \Gamma^L}{6m_\Phi^2} F_7\left(x_D\right)\,, \:\:
C_8=\ -{\cal N} \dfrac{ \Gamma^L}{2m_\Phi^2} F_7\left(x_D\right)\,,\\
C_{9}^{\prime \text{box}\,} &=& C_{9}^{\text{box}\,} \left(L\leftrightarrow R \right)\,,\qquad \qquad 
C_{10}^{\prime \text{box}\,}=-C_{10}^{\text{box}\,} \left(L\leftrightarrow R \right)\,, \\
C_{9}^{\prime\gamma}&=&C_{9}^{\gamma}\left(L\leftrightarrow R \right)\,,\qquad \qquad \quad \,\
C_{7,8}^{\prime}=C_{7,8}\left(L\leftrightarrow R \right)\,. \label{eq:C789p_4th}\
\eea
\medskip
\noindent $\bullet$ $b \to s \nu \bar\nu$ (see Eqs.~(\ref{eq:CL22_gen})-(\ref{eq:CR22_gen}))
\beq
C_L =
- {\cal N}\frac{\Gamma^L |\Gamma^L_\mu|^2}{{32\pi \alpha_{\text{EM}} m_{\Phi}^2}}
F(x_D,x_E) \,, \qquad
C_R =
- {\cal N}\frac{\Gamma^R |\Gamma^L_\mu|^2}{{32\pi \alpha_{\text{EM}} m_{\Phi}^2}}
F(x_D,x_E) \,.
\eeq

\noindent $\bullet$ $B_s-\bar{B_s}$ (see Eqs.~(\ref{eq:CBB_gen_beg})-(\ref{eq:CBB_gen_end}))
\beq
C_1 = \frac{|\Gamma^L |^2}{128 \pi^2 m_{\Phi}^2} F(x_D)  \,,
\quad
C_5 = - \frac{\Gamma^L \Gamma^R }{32 \pi^2 m_{\Phi}^2} F(x_D) \,,
\quad
\widetilde{C}_1 = \frac{|\Gamma^R|^2}{128 \pi^2 m_{\Phi}^2} F(x_D) \,,
\eeq
where the (simplified) loop function are defined in Appendix~\ref{app:loopfunctions}. In addition there are contributions to the $C_1$ analogue in $D^0-\bar{D^0}$ mixing obtained by substituting $L^b_1 \to L_1^c$ and $L^s_1 \to L_1^u$ within $\Gamma^L$.

In the charged-lepton sector $SU(2)$ breaking effects (encoded in $\lambda^E_{L,R}$) can give a sizable chiral enhancement of the NP effect in $a_\mu$ (see Eq.~\eqref{eq:amu_gen}) such that the long-standing anomaly in this channel can be addressed. In general one can parametrize the rotation matrices as
\beq
W^{E_{L,R}} = 
\begin{pmatrix}
 \cos(\theta_{L,R}) & -\sin(\theta_{L,R}) \\
 \sin(\theta_{L,R}) &  \cos(\theta_{L,R})
\end{pmatrix}\,,
\eeq
leading to
\bea
L_1^\mu &=& \Gamma_\mu^L \cos\theta_L \,, \qquad 
L_2^\mu = -\Gamma_\mu^L \sin \theta_L \,, \qquad 
L^\nu = \Gamma_\mu^L\,, \nonumber\\
R_1^\mu &=& \Gamma_\mu^R \sin\theta_R \,, \qquad 
R_2^\mu = \Gamma_\mu^R \cos\theta_R\,.
\eea
In our analysis we will consider a simplified setup with $\lambda^E_R = - \lambda^E_L \equiv \lambda^E$ that maximizes the effect in $a_\mu$ (which at leading order in $v$ is proportional to $\lambda^E_R-\lambda^E_L$). In this approximation we have for
\medskip

\noindent $\bullet$ $a_\mu$  (see Eq.~(\ref{eq:amu_gen}))
\bea\label{eq:amu_pheno}
\Delta a_\mu&=& 
\dfrac{ m_\mu^2}{8 \pi^2 m_{\Phi}^2} \left [
\left( |\Gamma^L_\mu|^2 + |\Gamma^R_\mu|^2 \right)
F_7\left(x_E\right) + \frac{8}{\sqrt{2}} \frac{v\,\lambda^E }{m_\mu}
\Gamma^{L}_\mu\Gamma^R_\mu 
G_7\left(x_E\right)
\right] \,,
\eea
where we have assumed real values for the couplings, implying a vanishing $d_{\mu}$.
Let us stress that the contributions proportional to $v \lambda^E$, coming from $SU(2)$ breaking terms, is chirally enhanced can give a sizable effect that can explain the $a_\mu$ anomaly.  
\medskip

\noindent $\bullet$ $Z\to\mu^+\mu^-$ (see Eqs.~(\ref{eq:gLZff_gen})-(\ref{eq:gRZff_gen}))
\bea
\Delta g_{\mu_L}(m_Z^2) &=& 
-\frac{|\Gamma^L_\mu|^2}{32\pi^2} \left[\dfrac{m_Z^2}{m_\Phi^2}
\left( (1-2 s^2_W)\widetilde G_9\!\left(x_E \right) 
+ \frac{2}{3} \left(\frac{v\lambda^E}{m_E} \right)^2 
F_{9}(x_E)
\right)  + \left(\frac{v\lambda^E}{m_E} \right)^2 F_{Z}(x_E)
\right]\,,\nonumber\\ \\
\Delta g_{\mu_R}(m_Z^2) &=& 
\frac{|\Gamma^R_\mu|^2}{32\pi^2} \left[\dfrac{m_Z^2}{m_\Phi^2}
\left(2 s^2_W \widetilde G_9\!\left(x_E \right)
+ \frac{2}{3}  \left(\frac{v\lambda^E}{m_E} \right)^2 F_{9}(x_E)
\right) + \left(\frac{v\,\lambda^E}{m_E} \right)^2 F_{Z}(x_E)
\right]\,, 
\eea
where the simplified loop function $F_{Z}(x_E)$ has been defined in Appendix~\ref{app:loopfunctions}. The results for $Z\to b\bar{b}$ couplings can be easily obtained by suitable substitutions. Note that in our approximation of $\lambda_{L,R}^D=0$ the correction to the $Z\bar sb$ vertex vanishes at $q^2=0$. Note that the UV divergences cancel as required, once for the couplings in \eq{eq:LZff} the relations
\beq
g^{\Psi,L(R)}=W^{E_{L(R)}\,\dag}\begin{pmatrix}
	g_{\Psi_{L,2}} 
	& 0 \\
	0 & g_{\Psi_e} 
\end{pmatrix}
W^{E_{L(R)}}
=W^{E_{L(R)}\,\dag}\begin{pmatrix}
	g_{\mu_L}^{\mathrm{SM}} 
	& 0 \\
	0 & g_{\mu_R}^{\mathrm{SM}}
\end{pmatrix}
W^{E_{L(R)}}\,,
\eeq
and $g_\Phi=0$ are used. Thus the finiteness of the result can be traced by to the unitarity of the matrices $W$.
\medskip

\subsection{Phenomenology}
\label{sec:pheno}

We are now ready to consider the phenomenology of our $4^{\rm th}$ generation model. For this purpose we will perform a combined fit to all the relevant and available experimental data, as briefly reviewed in Sec.~\ref{sec:Exp}. We perform this fit using the publicly available \texttt{HEPfit} package~\cite{deBlas:2019okz}, performing a Markov Chain Monte Carlo (MCMC) analysis employing the Bayesian Analysis Toolkit (BAT)~\cite{Caldwell:2008fw}.

Let us first choose specific values for the masses of the scalar $\Phi$ and the fermions $\Psi$. As observed in Ref.~\cite{Grinstein:2018fgb} a large splitting between the scalar mass and the vector-like lepton mass with respect to the vector-like quark masses is welcome to suppress the relative effect in $\Delta m_{B_s}$. Since the vector-like quarks should not be too light anyway because of direct LHC searches~\cite{Aaboud:2017wqg,Aaboud:2017vwy} we choose $m_\Phi \simeq m_E \simeq 450 \GeV$\footnote{Nearly degenerate masses $m_\Phi \simeq m_E$ are also welcome in the light of the dark matter relic density since the stable $\Phi$ is a suitable DM candidate. In fact, for $m_\Phi=450\GeV$, $450\le m_E \le 520\GeV$ the model allows for an efficient annihilation such that one does not over-shoot the matter density of the universe for order one $\Gamma$ couplings.} and $m_D=3.15 \TeV$, corresponding to $x_{E,L} \simeq 1$ and $x_D \simeq 50$. These values are well beyond the reach of direct searches at LHC: Concerning $m_{E,L}$ the bounds come from Drell-Yan production of the new fermions which are subsequently decaying in the neutral scalar and SM leptons. Therefore, the collider signature is similar to the one of MSSM slepton~\cite{Aaboud:2018jiw,CMS:2017fij}\footnote{A detailed study recasting these MSSM analysis for our model has been performed in Refs.~\cite{Kowalska:2017iqv,Calibbi:2018rzv}, finding $m_E \gtrsim m_\Phi =450 \GeV$ as an allowed solution.}. 

Turning to the coupling of the new scalars and fermions to quarks and muons, we assume a flatly distributed priors within the range $|\Gamma| \leq 1.5$ such that perturbativity is respected. The marginalized posterior probability  distribution for all NP couplings, together with the correlations among them, can be found in Appendix~\ref{app:posteriors}. In the rest of this section we will focus on one particular benchmark point, that we selected because it lies within all the 1$\sigma$ regions of the combined posterior distributions for the NP couplings (see Fig.~\ref{fig:posteriors}). The benchmark values are
\beq\label{eq:NPbenchmark}
|\Gamma_\mu^L| = 1.5\,, \qquad |\Gamma_\mu^R| = 1.4\,, \qquad \lambda^E = 0.0015\,, \qquad \Gamma^L = -1.0\,, \qquad \Gamma^R = -0.12\,
\eeq
assuming real values for all couplings. Note that the small value for $\lambda^E$ is obtained from the fit due to its correlation with $|\Gamma_\mu^R|$. As can be seen from the combined posterior distribution of these 2 parameters shown in Fig.~\ref{fig:posteriors}, higher values of $\lambda^E$ would require lower values of $|\Gamma_\mu^R|$, which is disfavored by the current fit to $b\to s\ell^+\ell^-$ data.

\begin{figure}[!t]
\centering
\subfigure{\includegraphics[width=.49\textwidth]{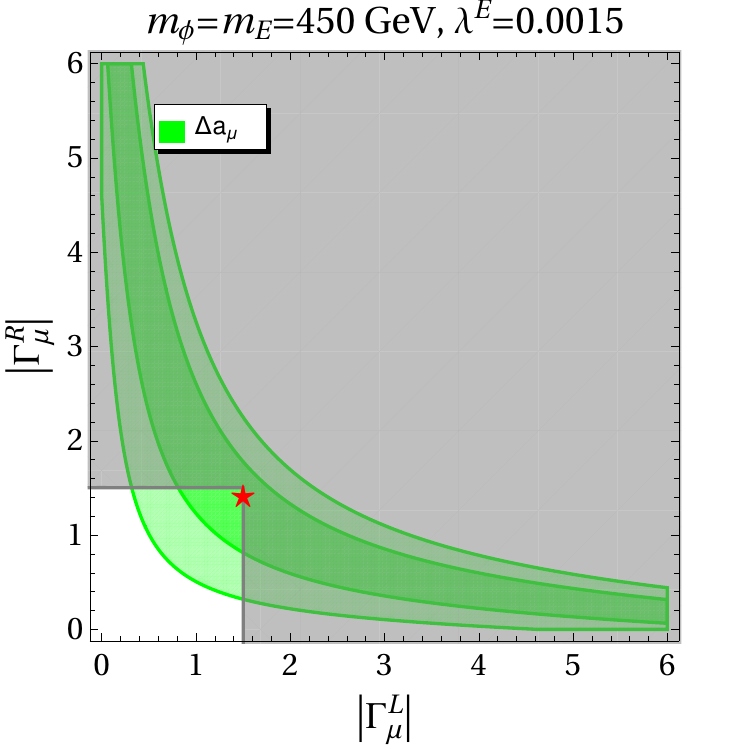}}\hspace{0.1cm}
\subfigure{\includegraphics[width=.49\textwidth]{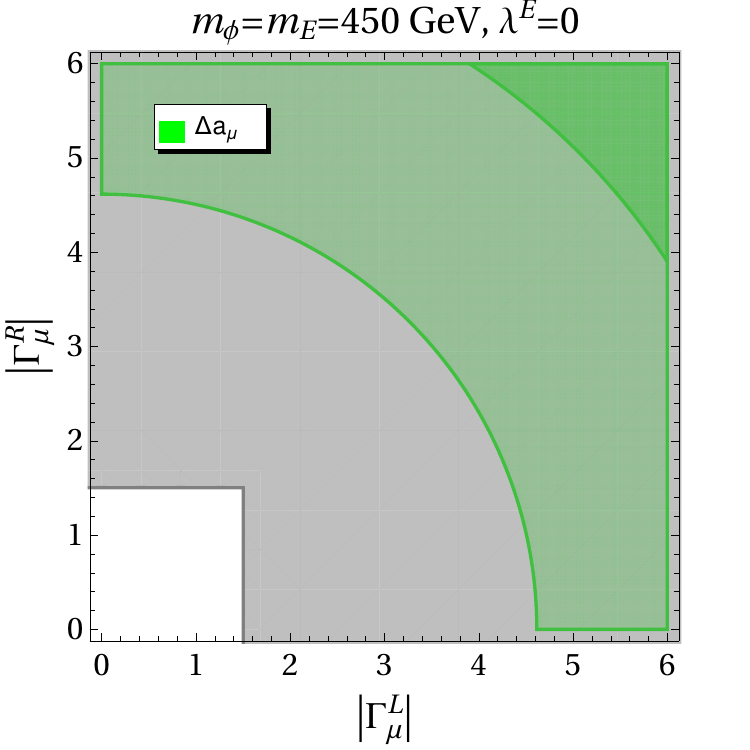}}
\caption{Left panel: allowed region for the coupling strength to the muon $|\Gamma_\mu^L|$ from the muon anomalous 
magnetic moment as a function of $|\Gamma_\mu^R|$, assuming $m_\phi = 450 \GeV$, $x^E$=1 and $\lambda^E = 0.0015$. The excluded region due to the requirement of perturbativity for $|\Gamma_\mu^{L,R}|$ is given in gray.
Dark (light) green corresponds to $1\sigma$ ($2 \sigma$) region. The red star marks the benchmark point $(1.5, 1.4)$. Right panel:  same as the left panel, but assuming $\lambda^E=0$.
\label{fig:gm2}}
\end{figure}

We observe that it is extremely important to allow for a right-handed coupling $\Gamma^R_\mu$ together a mixing coupling $\lambda^E$ in the muon sector such that $a_\mu$ can be explained. This can be seen from the fit in the $(|\Gamma_\mu^L|,|\Gamma_\mu^R|)$-plane from the left panel of Fig.~\ref{fig:gm2}. In the case with $\lambda^E=0.0015$, corresponding to the benchmark point reported in \eq{eq:NPbenchmark}, one can see that it is possible to explain the deviation in $a_\mu$ by means of couplings of order unity. However, the situation changes significantly if one did not allow the presence of a coupling of the vector-like leptons to the SM Higgs. As shown in the right panel of Fig.~\ref{fig:gm2}, with $\lambda^E=0$, it is not possible to obtain couplings that are perturbative and capable to give a satisfactory explanation of the anomalous magnetic moment of the muon at the same time. The presence of $\Gamma^R_\mu$ ameliorates the tension, but it is still not sufficient by itself to address the anomaly.

\begin{figure}[!t]
\centering
\subfigure{\includegraphics[width=.49\textwidth]{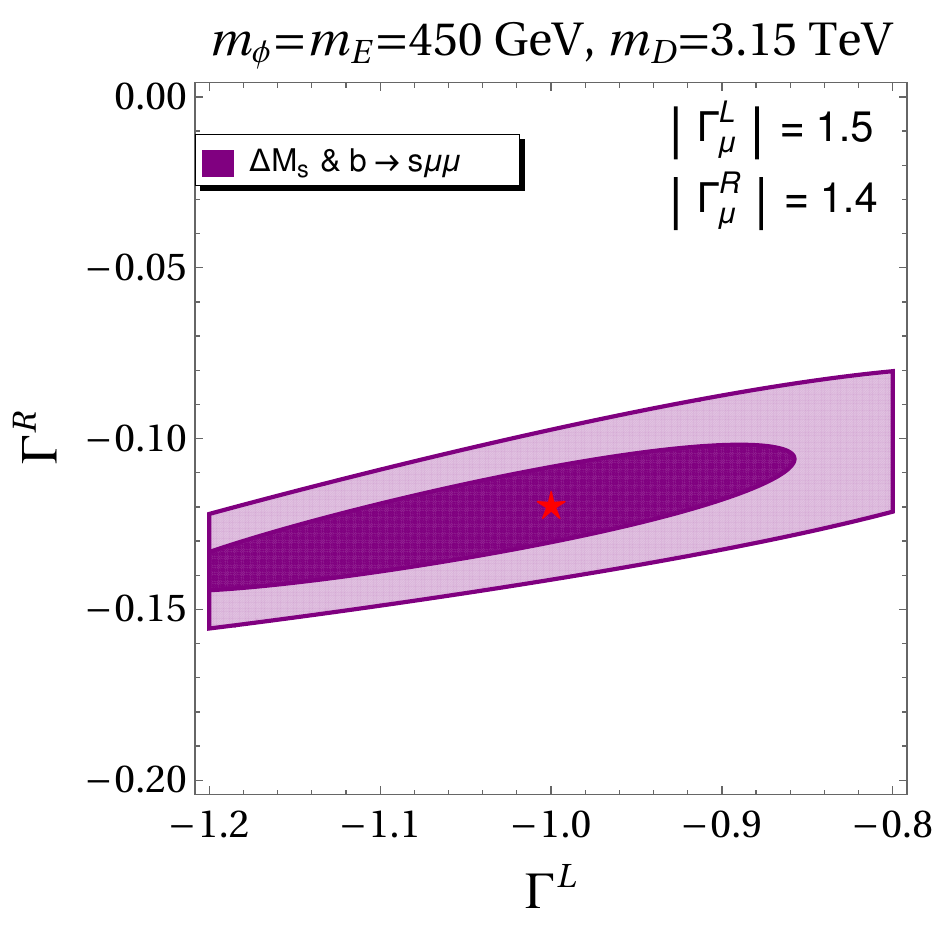}}
\subfigure{\includegraphics[width=.49\textwidth]{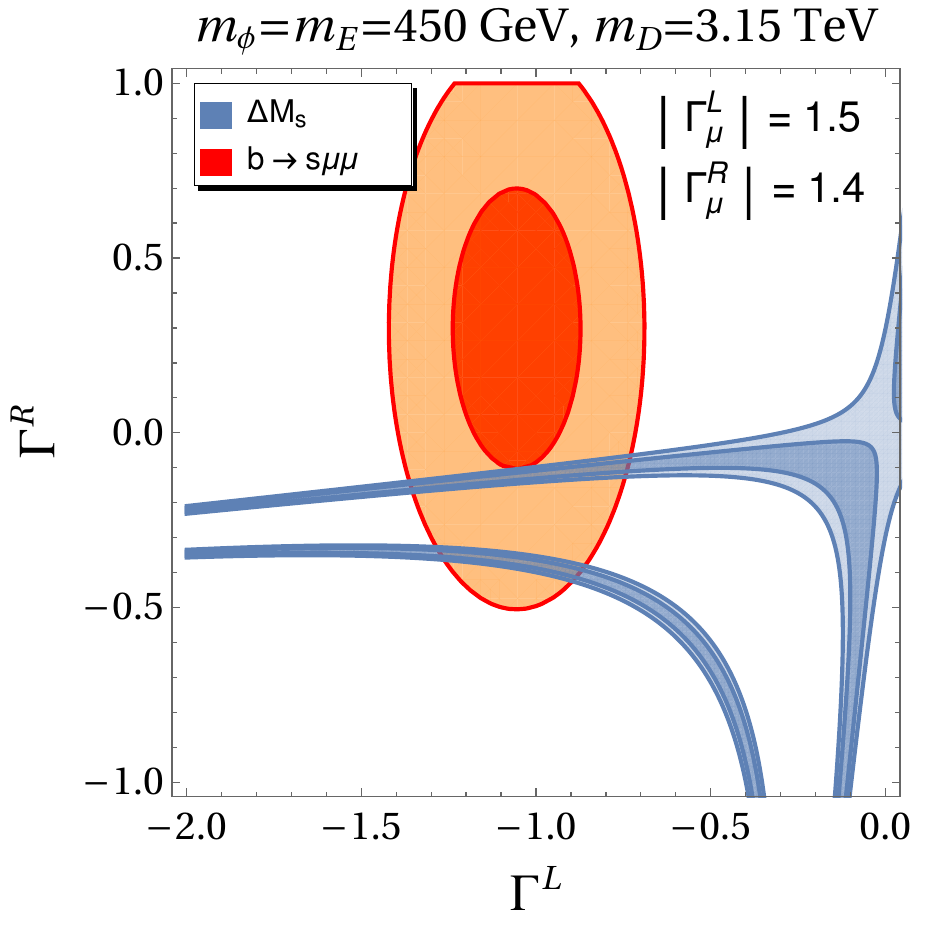}}
\hspace{0.1cm}
\caption{Left panel: Allowed region for the coupling $\Gamma^L\equiv 
	\Gamma_b^L \Gamma_s^{L*}$ and $\Gamma^R\equiv 
	\Gamma_b^R \Gamma_s^{R*}$ from $B_s-\bar{B_s}$ mixing and $b \to s\mu^+\mu^-$ data
for $m_\phi = m_E = 450 \,\text{GeV}$, $m_D = 3.15\,\text{TeV}$, $|\Gamma_\mu^L| = 1.5$ and $|\Gamma_\mu^R| = 1.4$. The red star marks the values at our benchmark point $(-1, -0.12)$.  The dark (light) purple regions is preferred at the $1\sigma$ ($2 \sigma$) level. Right panel: same as the left panel, but showing separately the allowed regions coming from $B_s-\bar{B_s}$ mixing (in blue) and $b \to s\mu^+\mu^-$ data (in red). The upper branch allowed by $B_s-\bar{B_s}$ mixing corresponds to the one shown on the left.
\label{fig:Bfit}}
\end{figure}

Also in the quark sector right-handed couplings are needed to address the $B$ anomalies without spoiling at the same time the measurement for $\Delta M_s$. This is particularly evident by looking at the left panel of Fig.~\ref{fig:Bfit}, where the region allowed by both $b \to s\mu^+\mu^-$ transitions and $B_s-{\bar B}_s$ is shown. Indeed, if one performs a separate fit to $b \to s\mu^+\mu^-$ transitions and $\Delta M_s$ as shown in the right panel of Fig.~\ref{fig:Bfit}, it is evident that the two channels are incompatible as long as one assumes a vanishing coupling to right-handed bottom and strange quarks, i.e. $\Gamma^R = 0$.

The preference for non-zero couplings (i.e. beyond the SM effects) is in general driven by $\Delta a_\mu$, the angular analyses of $B \to K^* \mu^+\mu^-$ and $B_s\to \phi\mu^+\mu^-$, the branching fraction of $B_s\to\mu^+\mu^-$ and the ratios $R_K$ and $R_{K^*}$. On the other hand, the experimental constraints coming from $b \to s\gamma$ and $B \to K^{(*)}\nu\bar\nu$ and $\Delta M_{B_s}$ set bounds on $\Gamma^{L,R}$ and $|\Gamma_\mu^L|$ that are less stringent than the ones obtained by the inclusion of the aforementioned channels involving $b \to s$ transitions in our setup with $\lambda^D_{L,R}=0$. Analogously, the constraints from $Z\to\mu^+\mu^-$ are found to give negligible constraints on $|\Gamma_{\mu}^{L,R}|$. Concerning $D_0-\bar{D_0}$ mixing, we recall that Eq.~\eqref{eq:LR_4th} implies a relation between $\Gamma^L\equiv L^b_1 L_1^{s*}$ and $L_1^{u,c}$. Exploiting the fact that only the product of $L_1^bL_1^{s*}$ enters $b\to s\ell^+\ell^-$, together with the suppression of the $L_1^b$ in $L_1^{u,c}$ by small CKM factors ($\mathcal{O}(\lambda^3)$ and $\mathcal{O}(\lambda^2)$, respectively), it is possible arrange the contributions to $\Gamma^L$ in such a way that the constraint imposed by $D_0-\bar{D_0}$ mixing is automatically satisfied. 
\medskip

We conclude this section by giving the results for some important observables (within our model) obtained from the global fit, namely
\bea
R_K[1.1,6] &=& 0.781(45)\,,\quad R_{K^*}[1.1,6] = 0.885(39)\,, \quad \mathcal{\bar{B}}(B_s\to \mu^+\mu^-) = 3.30(21)\cdot 10^{-9}\,, \nonumber\\ 
P_5'[4,6] &=& -0.454(69)\,, \qquad\quad P_5'[6,8] = -0.626(59)\,, \nonumber\\
\Delta a_\mu &=& 235(87)\cdot 10^{-11} \,, \qquad R_{\Delta M_s} = -0.02(8)\,.
\eea
All the predictions for these observables are compatible at the 1$\sigma$ level with their experimental measurements described in Sec.~\ref{sec:Exp}, except for $R_{K^*}$ which is compatible only at the $\sim 2 \sigma$ level. However, this is expected both from the global fit and from our specific model: As can be seen from the right panel of Fig.~\ref{fig:Bfit}, there is no overlap of the 1$\sigma$ regions from $b\to s \mu^+\mu^-$ data and $\Delta M_s$ in our model. Furthermore, since our model only allows for NP in muons (neglecting $Z,\gamma$ penguin effects), some small tensions are generated since flavour conserving $b\to s\mu^+\mu^-$ data prefers a smaller value of $R_K$ than the one recently measured.


\section{Conclusions and Outlook}
\label{sec:conclusions}

In this article we have studied in details the possibility that the intriguing anomalies in $b\to s\ell^+\ell^-$ processes are explained via box diagrams involving new scalars and fermions. Within this setup we have generalized previous analysis~\cite{Gripaios:2015gra,Arnan:2016cpy,Grinstein:2018fgb} to include couplings of the new particles to right-handed SM fermion and calculated the completely general expressions for the Wilson coefficients governing $b\to s$ processes ($b\to s\ell^+\ell^-$, $b\to s\nu\bar\nu$, $b\to s\gamma$ and $B_s - \bar{B}_s$ mixing). In addition, we have computed the effects in $a_\mu$ and $Z\to\mu^+\mu^-$ which unavoidably arise in such scenarios. 
\medskip

Furthermore, we have proposed a UV complete model containing a $4^{\rm th}$ vector-like generation of fermions and a new scalar, which is capable of explaining $b\to s\ell^+\ell^-$ data and $a_\mu$. We applied the formula derived in our generic setup (see Sec.~\ref{sec:generic}) to this model, illustrating their usefulness. In the following phenomenological analysis of our $4^{\rm th}$ generation model (see Sec.~\ref{sec:pheno}) we came to the conclusion that the $b\to s\ell^+\ell^-$ anomalies and $a_\mu$ can be explained simultaneously. As a benchmark point which can achieve this, and is consistent with direct LHC searches, we have used  $450 \GeV$ for the vector-like fermions and for the new scalar and $m_D=3.15 \TeV$ for the vector like quarks (with order one couplings). We have observed that right-handed couplings in the muon sector allow to address the long-standing anomaly in $a_\mu$ without having to require too large lepton couplings, if one allows for interactions of the vector-like fermions with the SM Higgs. Interestingly, due to the new results from LHCb and BELLE~\cite{Aaij:2019wad,Abdesselam:2019wac} the global fit to $b\to s\ell^+\ell^-$ data now prefers non-zero right-handed couplings to quarks and leptons as well, justifying the importance of our generalization of previous analysis performed in this work.
\medskip

Our $4^{\rm th}$ generation model is also interesting since $\Phi$ is a viable (stable) Dark Matter candidate. We briefly showed that if the mass of $\Phi$ is close to the one of the vector-like leptons, the correct relic density can be obtained while respecting the limits from Dark Matter direct detection. However, a more detailed investigation in the future seems worthwhile. 

We conclude observing that our formalism can be directly applied to $b\to d\ell^+\ell^-$ transitions. In fact, it leads to correlated effects in Kaon physics~\cite{Crivellin:2016vjc} if one aims at explaining the slight tensions in  $B\to \pi\mu^+\mu^-$~\cite{Hambrock:2015wka} simultaneously with the ones in $b\to s\ell^+\ell^-$ data. Furthermore, our setup and results can also be used for addressing the tension between theory and experiment in $\epsilon^\prime/\epsilon$~\cite{Buras:2015yba,Buras:2018ozh}\footnote{Note that calculations using chiral perturbation theory~\cite{Pallante:1999qf,Cirigliano:2011ny} instead are consistent with both the experimental measurement and the SM results of Refs.~\cite{Buras:2015yba,Buras:2018ozh}.}. Here, as a special case of our generic approach, the MSSM has already been studied with the conclusion that it can provide a valid explanation of the anomaly~\cite{Kitahara:2016otd,Endo:2016aws,Crivellin:2017gks}.
\bigskip

\noindent
\textbf{Note added:} After publication of this article, the g-2 collaboration at Fermilab released a new measurement of $a_\mu$~\cite{Abi:2021gix}, which together with the theory consensus of Ref.~\cite{Aoyama:2020ynm} leads to a tension of $4.2\,\sigma$. Furthermore, the LHCb collaboration updated the measurements of $R(K)$~\cite{Aaij:2021vac} and released the first measurement of $P_5^\prime$ in $B^+\to K^{+*}\mu\mu$~\cite{Aaij:2020ruw}. Note that this does not change the conclusions of our article and the impact on the numeric is very small. However, this gives additional support to our $4^{\rm th}$ generation model which can account for $(g-2)_\mu$ and $B$-anomalies.

\medskip


\acknowledgments{We thank Lorenzo Calibbi and Mauro Valli for useful discussions during the completion of the manuscript. The work of A.C. is supported by an Ambizione Grant (PZ00P2\_154834) and a Professorship Grant (PP00P2\_176884) of the Swiss National Science Foundation. P.A., M.F. and F.M. acknowledge the financial support from MINECO grant FPA2016-76005-C2-1-P,  Maria de Maetzu program grant MDM-2014-0367 of ICCUB and 2017 SGR 929.}


\appendix

\section{Fierz Identities}\label{app:fierz}
Here we list the Fierz identities for spinors used in the computations. With $i,j,k$ and $l$ representing Dirac indices here we find
\bea
\left(\gamma_\mu P_{L,R}\right)_{ij}\left(\gamma_\mu P_{L,R}\right)_{kl}&=&-\left(\gamma_\mu P_{L,R}\right)_{il}\left(\gamma_\mu P_{L,R}\right)_{kj}\\
\left(\gamma_\mu P_{L,R}\right)_{ij}\left(\gamma_\mu P_{R,L}\right)_{kl}&=&2\left( P_{R,L}\right)_{il}\left( P_{L,R}\right)_{kj}\\
\left( P_{L,R}\right)_{ij}\left(P_{L,R}\right)_{kl}&=&\dfrac{1}{2}\left( P_{L,R}\right)_{il}\left(P_{L,R}\right)_{kj}+\dfrac{1}{8}\left(\sigma_{\mu \nu}\right)_{il}\left(\sigma_{\mu \nu}P_{L,R}\right)_{kj}\\
\left( P_{L,R}\right)_{ij}\left(P_{R,L}\right)_{kl}&=&\dfrac{1}{2}\left(\gamma_\mu P_{R,L}\right)_{il}\left(\gamma_\mu P_{L,R}\right)_{kj}\\
\left( \sigma_{\mu \nu}\right)_{ij}\left(\sigma_{\mu\nu}P_{L,R}\right)_{kl}&=&6\left(\gamma_\mu P_{L,R}\right)_{il}\left(\gamma_\mu P_{L,R}\right)_{kj}-\dfrac{1}{2}\left( \sigma_{\mu \nu}\right)_{il}\left(\sigma_{\mu\nu}P_{L,R}\right)_{kj}\,,
\eea
where $P_{L,R}=\left(1\mp\gamma_5\right)/2$ and $\sigma_{\mu\nu}=\frac{i}{2}\left[\gamma_\mu,\gamma_\nu \right]$. When dealing with diagrams with crossed fermion lines, one needs Fierz identities involving charge conjugation matrices. Here, exchanging the second and the third Dirac index we find
\bea
\left(\gamma_\mu P_{L,R}C \right)_{ij}\left(C\gamma_\mu P_{L,R}\right)_{kl}&=&-2\left(P_{R,L}\right)_{ik}\left( P_{L,R}\right)_{jl}\\
\left(\gamma_\mu P_{L,R}C \right)_{ij}\left(C\gamma_\mu P_{R,L}\right)_{kl}&=&-\left(\gamma_\mu P_{L,R}\right)_{ik}\left(\gamma_\mu P_{R,L}\right)_{jl}\\
\left( P_{L,R}C \right)_{ij}\left(CP_{L,R}\right)_{kl}&=&\dfrac{1}{2}\left( P_{L,R}\right)_{ik}\left(P_{L,R}\right)_{jl}-\dfrac{1}{8}\left(\sigma_{\mu \nu}\right)_{ik}\left(\sigma_{\mu \nu}P_{L,R}\right)_{jl}\\
\left( P_{L,R}C \right)_{ij}\left(C P_{R,L}\right)_{kl}&=&-\dfrac{1}{2}\left(\gamma_\mu P_{R,L}\right)_{ik}\left(\gamma_\mu P_{R,L}\right)_{jl}\,,
\eea
with the charge conjugation matrix defined as $C=i\gamma_0\gamma_2$.


\section{Loop Functions}\label{app:loopfunctions}

Here we list the dimensionless loop functions introduced in Sections~\ref{sec:generic} and~\ref{sec:4thgen}. The loop functions appearing in box diagrams that involve four different masses are defined as

\begin{equation}
\begin{aligned}
F(x,y,z) &= 
\frac{x^2 \log (x)}{(x-1) (x-y) (x-z)}+\frac{y^2 \log (y)}{(y-1) (y-x) (y-z)}+\frac{z^2 \log (z)}{(z-1) (z-x) (z-y)} \,,\\
G(x,y,z) &=2\left( \frac{x \log (x)}{(x-1) (x-y) (x-z)}+\frac{y \log (y)}{(y-1) (y-x) (y-z)}+\frac{z \log (z)}{(z-1) (z-x) (z-y)} \right)\,,
\end{aligned}
\label{eq:FG3}
\end{equation} 
which in the equal mass limit read
\begin{equation}
F(1,1,1)=-G(1,1,1)=\frac{1}{3}\,.
\end{equation} 
In the presence of only three different masses in the loop, one gets the functions
\begin{equation}
\begin{aligned}
F(x,y) &\equiv F(x,y,1) = 
\frac{1}{(1-x) (1-y)}+\frac{x^2 \log (x)}{(1-x)^2 (x-y)}+\frac{y^2 \log (y)}{(1-y)^2 (y-x)} \,,\\
G(x,y) &\equiv G(x,y,1) = 
2\left( \frac{1}{(1-x) (1-y)}+\frac{x \log (x)}{(1-x)^2 (x-y)}+\frac{y \log (y)}{(1-y)^2 (y-x)} \right)\,,
\end{aligned}
\label{eq:FG2}
\end{equation}
while, in the presence of only two different masses in the loop, one gets
\begin{equation}
\begin{aligned}
F(x) &\equiv F(x,x) = 
\frac{x+1}{(x-1)^2} - \dfrac{2 x \log (x)}{(x-1)^3} \,,\\
G(x) &\equiv G(x,x) =
\frac{2}{(x-1)^2} - \dfrac{(x+1) \log (x)}{(x-1)^3} \,.
\end{aligned}
\label{eq:FG1}
\end{equation}
The loop functions appearing in photon- and gluon-penguin diagrams are defined as
\begin{equation}
\begin{aligned}
F_7(x)&=\dfrac{ x^3-6 x^2+3 x+2+6 x \log x}{12 (x-1)^4}\,,
&\widetilde F_7(x)&=x^{-1} F_{7}(x^{-1})\,, \\
G_7(x)&=\dfrac{x^2  - 4x + 3+ 2 \log x}{8 (x-1)^3}\,, 
&\widetilde G_7(x)&=\frac{x^2-2 x\log x -1}{8 (x-1)^3}\,, \\
F_9(x)&=\dfrac{-2x^3+9x^2-18x+11+6 \log x}{36(x-1)^4}\,, 
&\widetilde F_9(x)&=x^{-1}F_9(x^{-1})\,, \\
G_9(x)&=\dfrac{-16x^3+45x^2 -36x + 7 +6(2x-3)x^2\log x}{36(x-1)^4}\,,
&\widetilde G_9(x)&=x^{-1} G_9(x^{-1})\,,
\label{eq:FG79}
\end{aligned}
\end{equation} 
which in the equal mass limit read
\begin{equation}
F_7(1)=\widetilde F_7(1)=\frac{G_7(1)}{2}=\widetilde G_7(1)=-F_9(1)=-\widetilde F_9(1)=\frac{G_9(1)}{3}=\frac{\widetilde G_9(1)}{3}=\frac{1}{24}\,.
\end{equation} 
Finally, the loop functions for the calculation of $Z$-penguins are defined as
\bea
G_Z(x,y)&=& x F_V(x,y)+ x\leftrightarrow y \,, \nonumber \\
F_Z(x,y,m)&\equiv& \overline{F}_Z(x,y) - \overline{div}_\varepsilon =  \left(x^2 F_V(x,y) + x\leftrightarrow y\right) - \overline{div}_\varepsilon \,, \nonumber \\
H_Z(x,y,m)&\equiv& \overline{H}_Z(x,y) + \overline{div}_\varepsilon = \left(y F_V(x,y) + x\leftrightarrow y\right) + 1 + \overline{div}_\varepsilon \,, \nonumber \\
I_Z(x,m)&\equiv& \overline{I}_Z(x) + \overline{div}_\varepsilon =\dfrac{x}{x-1}- x^2 F_V(x,1) + \overline{div}_\varepsilon \,, \nonumber
\\
\widetilde G_Z(x,y)&=&  x K_V(x,y)+ x\leftrightarrow y \,, \nonumber\\
\widetilde F_Z(x,y)&=&\left(
x^2 K_V(x,y)- \frac{x^2}{x-y}F_V(x,y) \right)+ x\leftrightarrow y \,, \nonumber\\
\widetilde H_Z(x,y)&=& \left(
\dfrac{x^2 y}{(y-1) (x-y)^2}-\dfrac{x^2 y^2 (3 x-y-2) \log (x)}{(x-1)^2 (x-y)^3} \right)+ x\leftrightarrow y \,, 
\eea
where we have defined $\overline{div}_\varepsilon = \Delta_\varepsilon-\log\left(\dfrac{m^2}{\mu^2}\right)$ and
\bea
F_V(x,y) = \frac{\log (x)}{(x-1) (x-y) }\,,\qquad
K_V(x,y) = 
\frac{\left(x^2+x y-2 y\right) \log (x)}{(x-1)^2 (x-y)^3}-\frac{1}{(x-1)(x-y)^2}\,.\nonumber\\
\eea
It is interesting to notice that the following relations hold between particular limits of the penguin induced functions:
\begin{flalign}\label{eq:ZLF_limits}
&H_Z\left(\frac{m^2}{n^2},\frac{m^2}{n^2},m\right) = I_Z\left(\frac{m^2}{n^2},n\right) \nonumber \,,&
\\
&- \frac{m^2}{n^2} G_Z\left(\frac{m^2}{n^2},\frac{m^2}{n^2}\right) + \frac12 F_Z\left(\frac{m^2}{n^2},\frac{m^2}{n^2},n\right) + \frac14 H_Z\left(\frac{m^2}{n^2},\frac{m^2}{n^2},m\right) + \frac14 I_Z\left(\frac{m^2}{n^2},n\right) = 0 \nonumber \,,&
\\
&\frac12 x\, \widetilde G_Z(x,x) - \frac13 \widetilde F_Z(x,x) = \widetilde G_9(x) \nonumber \,,&
\\
&
\frac{1}{6}\, \widetilde F_Z(x,x) = F_9(x) \,.&
\\
&
\frac{1}{6 x}\, \widetilde H_Z(x,x) = \widetilde F_9(x) \,.&
\end{flalign}
Moreover, it is useful to define the limit
\beq
F_Z(x) \equiv \overline{F}_Z(x,x)  = \frac{x}{x-1} + \frac{(x-2)x\log x}{(x-1)^2}\,.
\eeq
Finally, the equal mass limits read
\begin{equation}
G_Z(1,1)=\frac{\overline{F}_Z(1,1)}{3}=-\overline{H}_Z(1,1)=-\overline{I}_Z(1)=6\widetilde G_Z(1,1)=-2\widetilde F_Z(1,1)=-2\widetilde H_Z(1,1)=\frac{1}{2}\,.
\end{equation} 


\boldmath
\section{Real Scalars and Majorana Fermions }
\label{app:Crossedbsmumuboxes}
\unboldmath

\begin{figure}[!t]
	\centering
	\includegraphics[scale=0.56]{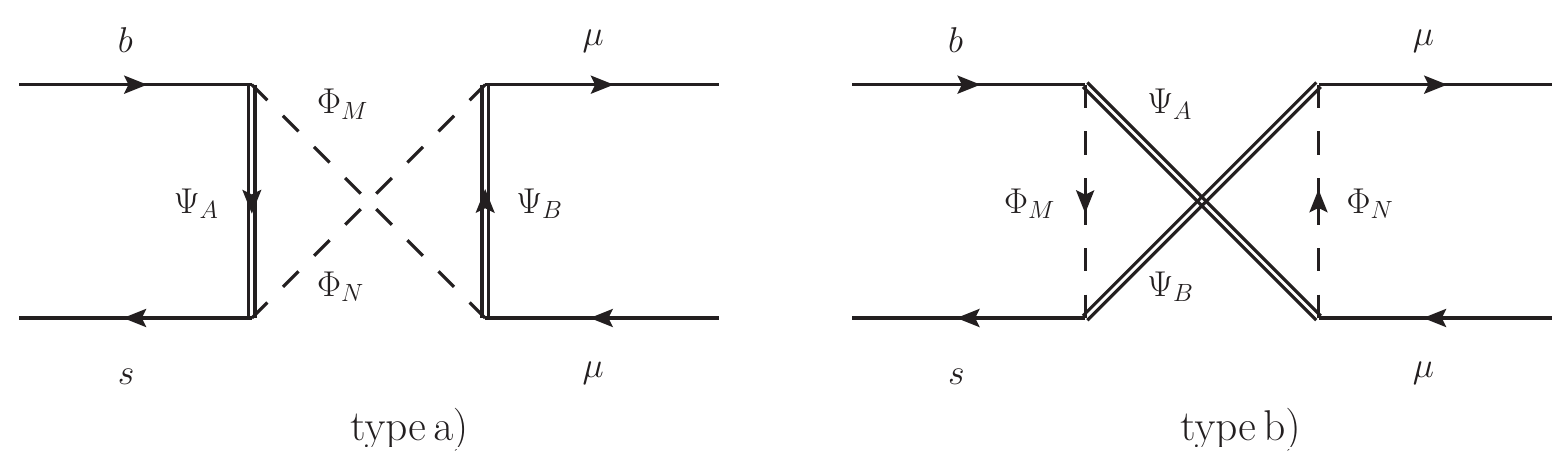}
	\caption{Crossed box diagrams contributing to $b\to s\mu^+\mu^-$ transitions. The diagram on the left appears in models with real scalars, while the one on the right can be constructed in models with Majorana fermions.}
	\label{fig:diagbsmumucrossed}
\end{figure}

\begin{figure}[!t]
	\centering
	\includegraphics[scale=0.56]{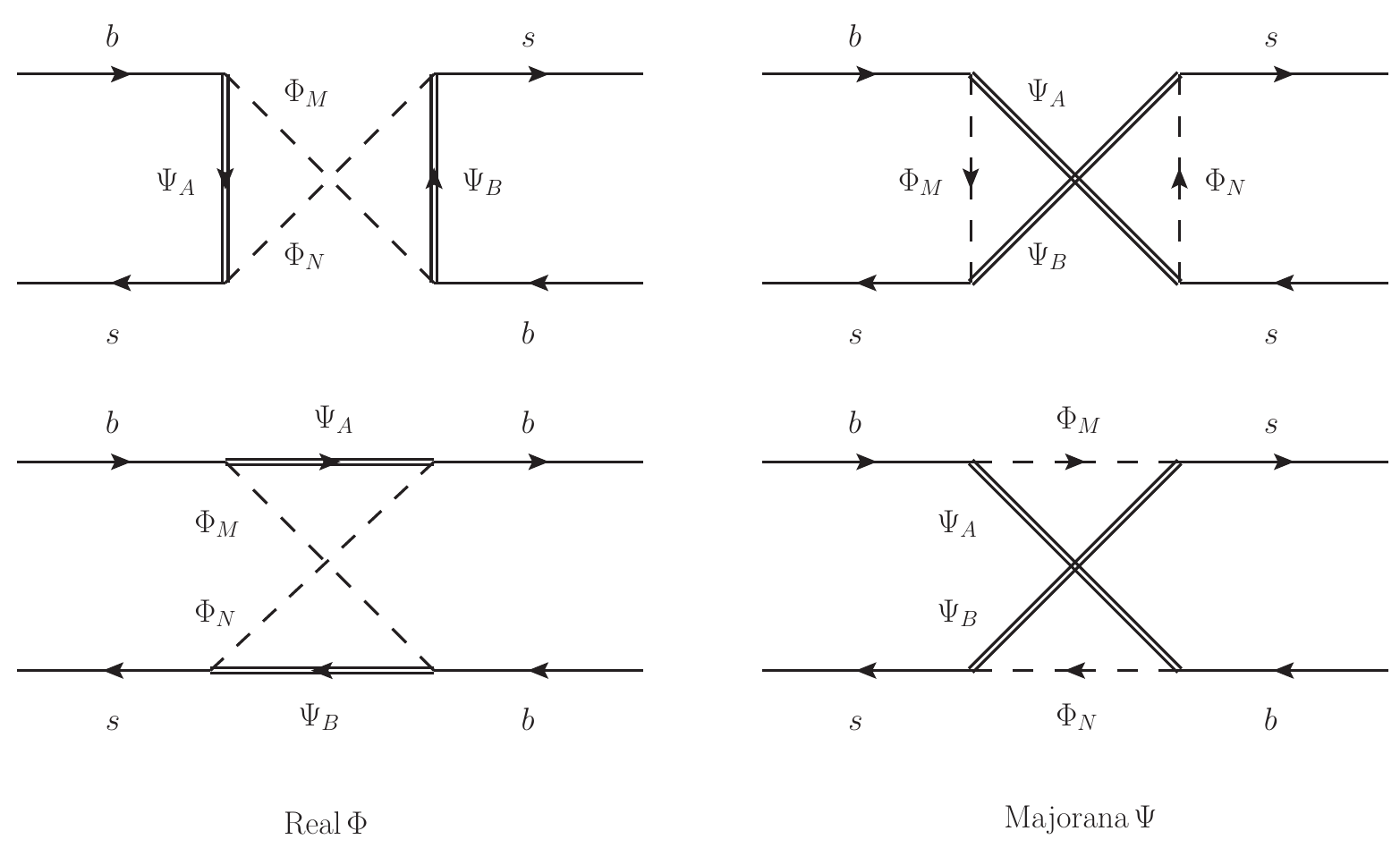}
	\caption{Box diagrams contributing to $B_s-\bar B_s$ mixing. The diagram on the left is relative to models with real scalars, while the one on the right refers to models with Majorana fermions.}
	\label{fig:diagBscrossed}
\end{figure}

If the NP fields have the appropriate quantum numbers they can be either real scalars or Majorana fermions. If this is the case, crossed diagrams as shown in Fig.~\ref{fig:diagbsmumucrossed} can be constructed and contribute to $b\to s \mu^+ \mu^-$ transitions in addition to the ones shown in Fig.~\ref{fig:diagbsmumu}. Similarly, there are contributions from crossed boxes to $B_s-\bar B_s$ mixing (in addition to the ones in Fig.~\ref{fig:diagBBmix}) arising due to the diagrams in Fig.~\ref{fig:diagBscrossed}.
\medskip

\begin{table}[!t]
	\centering
	\begin{tabular}{c | cccc | c | cccc | c}
		$SU(3)$, type $a)$ & $\Psi_A$ & $\Psi_B$ & $\Phi_M$ & $\Phi_N$ & 
		$SU(3)$, type $b)$ & $\Psi_A$ & $\Psi_B$ & $\Phi_M$ & $\Phi_N$ & $\chi$ \\
		\hline
		I   & 3   & 1   & 1   & 1     &   I   & 1   & 1   & 3   & 1     &   1 \\ 
		III   & 3   & 8   & 8   & 8     &   III   & 8   & 8   & 3   & 8     &   4/3 \\
	\end{tabular}
	\caption{Table of $SU(3)$-factors entering the box induced Wilson coefficients involved in $b \to s$ transitions for real scalars, type $a)$, and Majorana fermions, type $b)$. The numbers of each representation refer to the ones in Table~\ref{tab:SU3_bsmumu}.
		\label{tab:SU3_bsmumucrossed}}
\end{table} 

\boldmath
\subsection{\texorpdfstring{$b\to s \mu^+\mu^-$}{btosmu+mu-}}
\label{app:crossedbsmumu}
\unboldmath

In $b\to s \mu^+ \mu^-$ the possible representations that give rise to additional crossed diagrams with real scalars or Majorana fermions are listed in Tab.~\ref{tab:SU3_bsmumucrossed}. For type $a)$ the only possibility is to have real scalars, while for type $b)$ one can only have crossed diagrams in the presence of Majorana fermions.
\medskip

The contribution to the Wilson coefficients stemming from the diagrams in Fig.~\ref{fig:diagbsmumucrossed}$a)$ corresponds to the ones listed for $a)$-type in Eqs.~\eqref{eq:C9_gen}-\eqref{eq:CP_gen}, after inverting $M \leftrightarrow N$ in the muon couplings and changing $F(x_{AM},x_{BM},x_{NM})\to-F(x_{AM},x_{BM},x_{NM})$. For case $b)$ (see right diagram in Fig.~\ref{fig:diagbsmumucrossed}) the Wilson coefficients are given by 
\bea
C_9^{{\text{box}}\; b)} &=& - {\cal N}
\frac{\chi L^{s*}_{BM}L^b_{AM}}{{32\pi \alpha_{\text{EM}} m_{\Phi_M}^2}}
\bigg[ 
L^{\mu *}_{BN}L^\mu_{AN} \frac{m_{\Psi_A} m_{\Psi_B}}{m_{\Phi_M}^2}G(x_{AM},x_{BM},x_{NM})  \nonumber \\ 
&&  \qquad\qquad\qquad\qquad -
R^{\mu *}_{BN}R^\mu_{AN}F(x_{AM},x_{BM},x_{NM}) 
\bigg]\,,
\label{eq:C9_cross} \\
C_{10}^{{\text{box}}\; b)} &=& {\cal N}
\frac{\chi L^{s*}_{BM}L^b_{AM}}{{32\pi \alpha_{\text{EM}} m_{\Phi_M}^2}}
\bigg[ 
L^{\mu *}_{BN}L^\mu_{AN}\frac{m_{\Psi_A} m_{\Psi_B}}{m_{\Phi_M}^2}G(x_{AM},x_{BM},x_{NM})  \nonumber \\ 
&&  \qquad\qquad\qquad\quad
+
R^{\mu *}_{BN}R^\mu_{AN} 
F(x_{AM},x_{BM},x_{NM})\bigg] \,,
\label{eq:C10_cross} \\
C_S^{{\text{box}}\; b)} &=& {\cal N}
\frac{\chi L^{s*}_{BM}L^b_{AM}}{{16\pi \alpha_{\text{EM}} m_{\Phi_M}^2}}
\bigg[ 
R^{\mu *}_{BN}L^\mu_{AN}F(x_{AM},x_{BM},x_{NM}) \nonumber \\ 
&&  \qquad\qquad\qquad\quad
+
L^{\mu *}_{BN}R^\mu_{AN} \frac{m_{\Psi_A} m_{\Psi_B}}{2m_{\Phi_M}^2}G(x_{AM},x_{BM},x_{NM})
\bigg]\,,
\label{eq:CS_cross} \\
C_P^{{\text{box}}\; b)} &=& {\cal N}
\frac{\chi L^{s*}_{BM}L^b_{AM}}{{16\pi \alpha_{\text{EM}} m_{\Phi_M}^2}}
\bigg[ 
R^{\mu *}_{BN}L^\mu_{AN}F(x_{AM},x_{BM},x_{NM}) \nonumber \\ 
&& \qquad\qquad\qquad\quad
-
L^{\mu *}_{BN}R^\mu_{AN} \frac{m_{\Psi_A} m_{\Psi_B}}{2m_{\Phi_M}^2}G(x_{AM},x_{BM},x_{NM})
\bigg]\,,
\label{eq:CP_cross} \\
C_T^{{\text{box}}\; b)} &=& -{\cal N}
\frac{\chi
L^{s*}_{BM}R^b_{AM} L^{\mu *}_{BN}R^\mu_{AN}
}{{16\pi \alpha_{\text{EM}} m_{\Phi_M}^2}}
\frac{m_{\Psi_A} m_{\Psi_B}}{m_{\Phi_M}^2}  G(x_{AM},x_{BM},x_{NM})\,,
\label{eq:CT_cross}
\eea
\bea
C_{9,S}^{\prime \text{box}\,}=C_{9,S}^{\text{box}\,} \left(L\leftrightarrow R \right)\,,
\qquad C_{P,10}^{\prime \text{box}\,}=-C_{P,10}^{\text{box}\,} \left(L\leftrightarrow R \right)\,,
\label{eq:C910SPTp_cross}
\eea

\begin{table}[!t]
\centering
\begin{tabular}{c | cccc | cc l}
$SU(3)$ & $\Psi_A$ & $\Psi_B$ & $\Phi_M$ & $\Phi_N$ & $\chi_{BB}^M$ & $\tilde{\chi}_{BB}^M$& \\
\hline
I   & 3       & 3       & 1       & 1       &  1    &  0&Real $\Phi$\\  
II  & 1       & 1       & 3       & 3       &  0    &  1&Majorana $\Psi$\\ 
III & 3       & 3       & 8       & 8       &  5/18 &  -1/6&Real $\Phi$ \\
IV  & 8       & 8       & 3       & 3       &  -1/6 &  5/18&Majorana $\Psi$ \\ 
V   & 3       & 3       & (1,8)   & (8,1)   & 1/6  &  -1/2&Real $\Phi$  \\  
VI  & (1,8)   & (8,1)   & 3       & 3       &  -1/2  & 1/6&Majorana $\Psi$  \\  
\end{tabular}
\caption{Table of $SU(3)$-factors entering the box induced Wilson coefficients involved in $B_s-\bar{B}_s$ mixing for real scalars and Majorana fermions.
\label{tab:SU3_Bscrossed}}
\end{table}

\boldmath
\subsection{\texorpdfstring{$B_s-\bar{B_s}$}{Bs-barBs} mixing }
\label{app:crossedBs}
\unboldmath
For $B_s$ mixing we can either have real scalars or Majorana fermions. In Tab.~\ref{tab:SU3_Bscrossed} we list the possible representations of the diagrams in Fig.~\ref{fig:diagBscrossed} writing explicitly if we have a real scalar contribution (diagrams on the left side of the figure) or a Majorana fermion (diagrams on the right). The WCs for real scalar crossed diagrams correspond to the ones listed in Eqs.\textbf{~\eqref{eq:C9_gen}-\eqref{eq:CP_gen}}, after inverting $M \leftrightarrow N$ in two of the four couplings and changing $F(x_{AM},x_{BM},x_{NM})\to-F(x_{AM},x_{BM},x_{NM})$, whereas matching to the generic Lagrangian from Eq.~(\ref{eq:L_bsmumu}) with the crossed fermion contributions, one obtains the following results for the coefficients:
\bea
C_1&=&
(\chi_{BB}^M + \tilde \chi_{BB}^M)
\frac{L^{s*}_{AN}L^b_{BM} L^{s*}_{AM}L^b_{BN}}{128 \pi^2 m_{\Phi_M}^2} 
\frac{m_{\Psi_A} m_{\Psi_B}}{m_{\Phi_M}^2}G\hspace{-2.8pt}\left(x_{AM},x_{BM},x_{NM} \right) \,,	 \label{eq:CBB_cross_Ani}
\\
C_{2,3}&=& 
-(\chi_{BB}^M + \tilde \chi_{BB}^M)
\frac{R^{s*}_{AN}L^b_{BM} R^{s*}_{AM}L^b_{BN}}{64 \pi^2 m_{\Phi_M}^2} 
\frac{m_{\Psi_A} m_{\Psi_B}}{m_{\Phi_M}^2}G\hspace{-2.8pt}\left(x_{AM},x_{BM},x_{NM} \right) \,,
\\
C_4&=&
\frac{L^b_{AM}R^b_{AN}}{32 \pi^2 m_{\Phi_M}^2} 
\left[  
  \chi_{BB}^M L^{s*}_{BM}R^{s*}_{BN}   -
\tilde \chi_{BB}^M R^{s*}_{BM}L^{s*}_{BN}
\right]F\hspace{-2.8pt}\left(x_{AM},x_{BM},x_{NM} \right)\,,\qquad
 \\
C_5&=&
\frac{L^b_{AM}R^b_{AN}}{32 \pi^2 m_{\Phi_M}^2} 
\left[  
 \tilde\chi_{BB}^M  L^{s*}_{BM}R^{s*}_{BN} -
\chi_{BB}^M R^{s*}_{BM}L^{s*}_{BN}
\right]F\hspace{-2.8pt}\left(x_{AM},x_{BM},x_{NM} \right)\,,\qquad
  \\
  \widetilde C_i &=& C_i \left(L \to R \right)\,, \qquad \text{for} \qquad i = \{1,2,3\}\,,
  \label{eq:CBB_majorana}
\eea
The corresponding contributions to $D_0-\overline{D}_0$ mixing are obtained from Eqs.~(\ref{eq:OPBB})-(\ref{eq:CBB_gen_end}) via the replacements $s \to u$ and $b \to c$.
\medskip

\boldmath
\section{Crossed Diagrams with Complex Scalars  }
\label{app:Crossedbsmumucomplex}
\unboldmath

\begin{figure}[!t]
\centering
\includegraphics[scale=0.56]{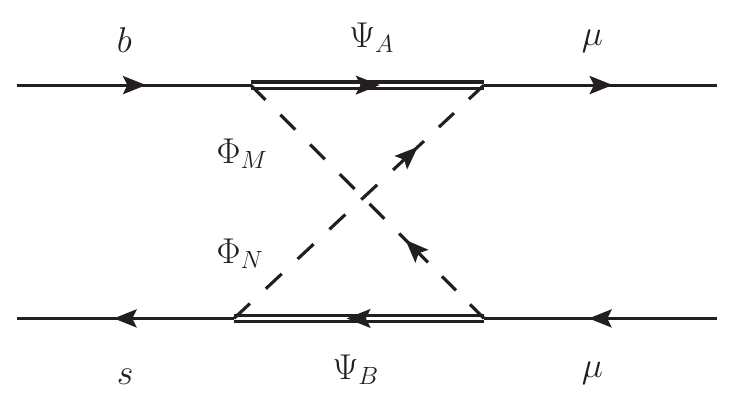}
\caption{Crossed box diagrams contributing to $b\to s\mu^+\mu^-$ transitions. The diagram appears when a complex scalar couples to $b,s$ quarks and its conjugate couples to the muons.}
\label{fig:diagbsmumucrossedcomplex}
\end{figure}

There is also the possibility that a complex scalar couples to the down-type quarks whereas its hermitian conjugated version couples to muons. This means that the Lagrangian in Eq.~\eqref{eq:L_bsmumu} takes a slightly different form, namely
\begin{eqnarray} \label{eq:L_bsmumu_complexcrossedscalars}
\mathcal{L}_{\rm int} = & \bigg[ & \bar \Psi _A\left( L_{AM}^b{P_L}{b } + L_{AM}^s{P_L}{s }
+ {R_{AM}^b}{P_R}{b } + R_{AM}^s{P_R}{s }  \right){\Phi _M}  \nonumber \\
&&+\bar{\Psi} _A\left(  L_{AM}^\mu {P_L} +  {R_{AM}^\mu {P_R}\mu } \right){\Phi _M^\dag} \bigg] + \hc\, .
\end{eqnarray}
Also this Lagrangian generates a contribution to $b\to s \mu^+ \mu^-$ via the diagram shown in Fig.~\ref{fig:diagbsmumucrossedcomplex}. The possible representations under the $SU(3)$ of the new scalars and fermions in the loop are listed in Tab.~\ref{tab:SU3_bsmumucrossedcomplex}. The corresponding Wilson Coefficients can be obtained from the ones calculated for the type $b)$ diagrams in Eqs.~\eqref{eq:C9_gen}-\eqref{eq:CP_gen}  by exchanging  
$M\leftrightarrow N$ in the couplings $R,L$ and replacing
$F(x_{AM},x_{BM},x_{NM})\rightarrow-F(x_{AM},x_{BM},x_{NM})$.
\begin{table}[!t]
\centering
\begin{tabular}{c | cccc | c}
$SU(3)$& $\Psi_A$ & $\Psi_B$ & $\Phi_M$ & $\Phi_N$&$\chi$ \\
\hline
I   & (1,3)   & (3,1)   & ($\bar{3}$,1)   & (1,$\bar{3}$) & 1 \\ 
II   & (8,3)   & (3,8)   & ($\bar{3}$,8)   & (8,$\bar{3}$) & 4/3 \\ 
III   & $\bar{3}$   & $\bar{3}$  &3   & 3   & 2 \\
\end{tabular}
\caption{Table of $SU(3)$-factors entering the box induced Wilson coefficients involved in $b \to s$ transitions for crossed diagrams with complex scalars.
\label{tab:SU3_bsmumucrossedcomplex}}
\end{table}


\section{Posterior Distributions}\label{app:posteriors}
\begin{figure}[!t]
\hspace{2.65em}\subfigure{\includegraphics[width=.174\textwidth]{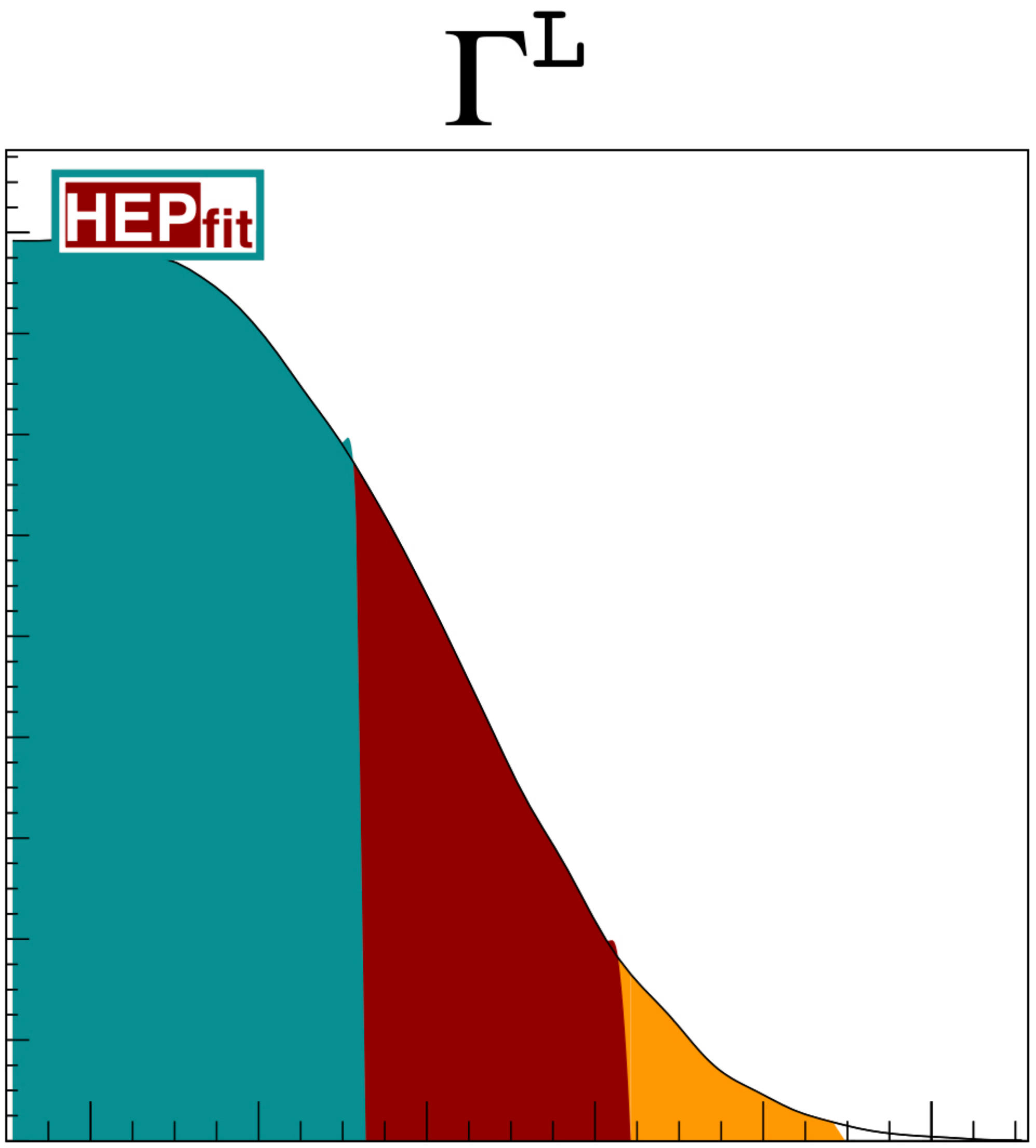}}
\vspace{-1.7em} \\
\subfigure{\includegraphics[width=.24\textwidth]{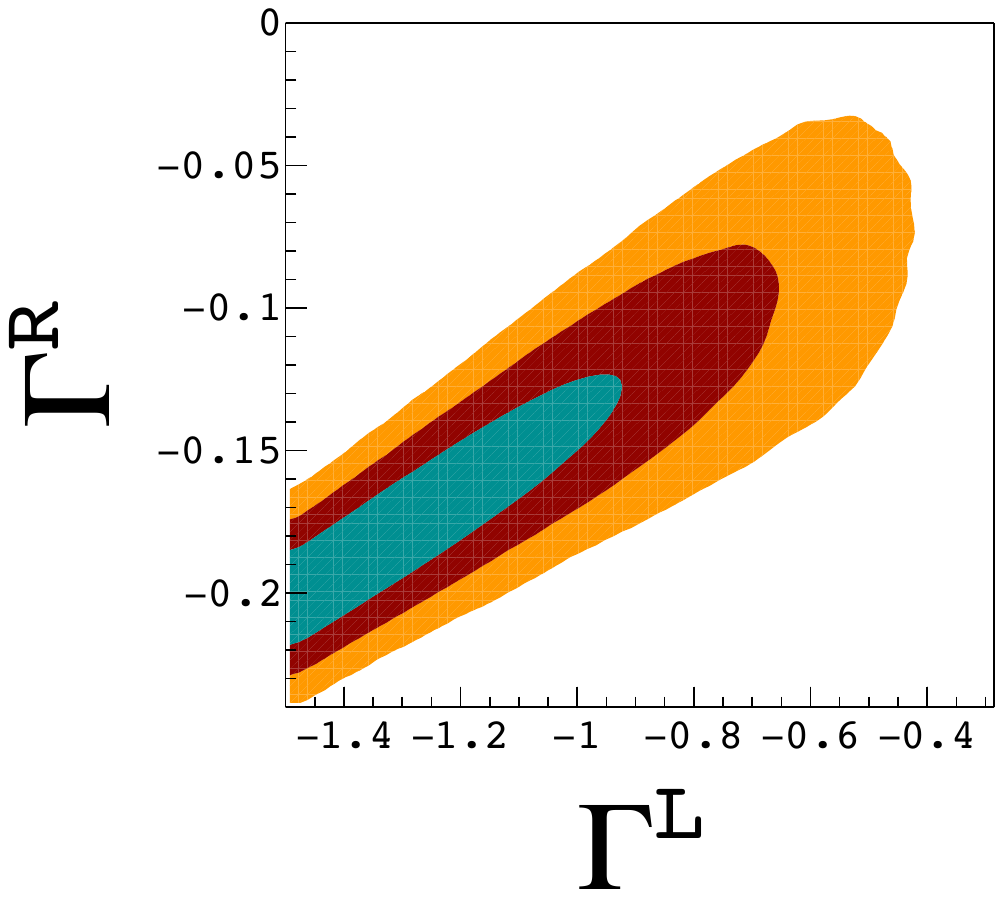}}
\subfigure{\includegraphics[width=.168\textwidth]{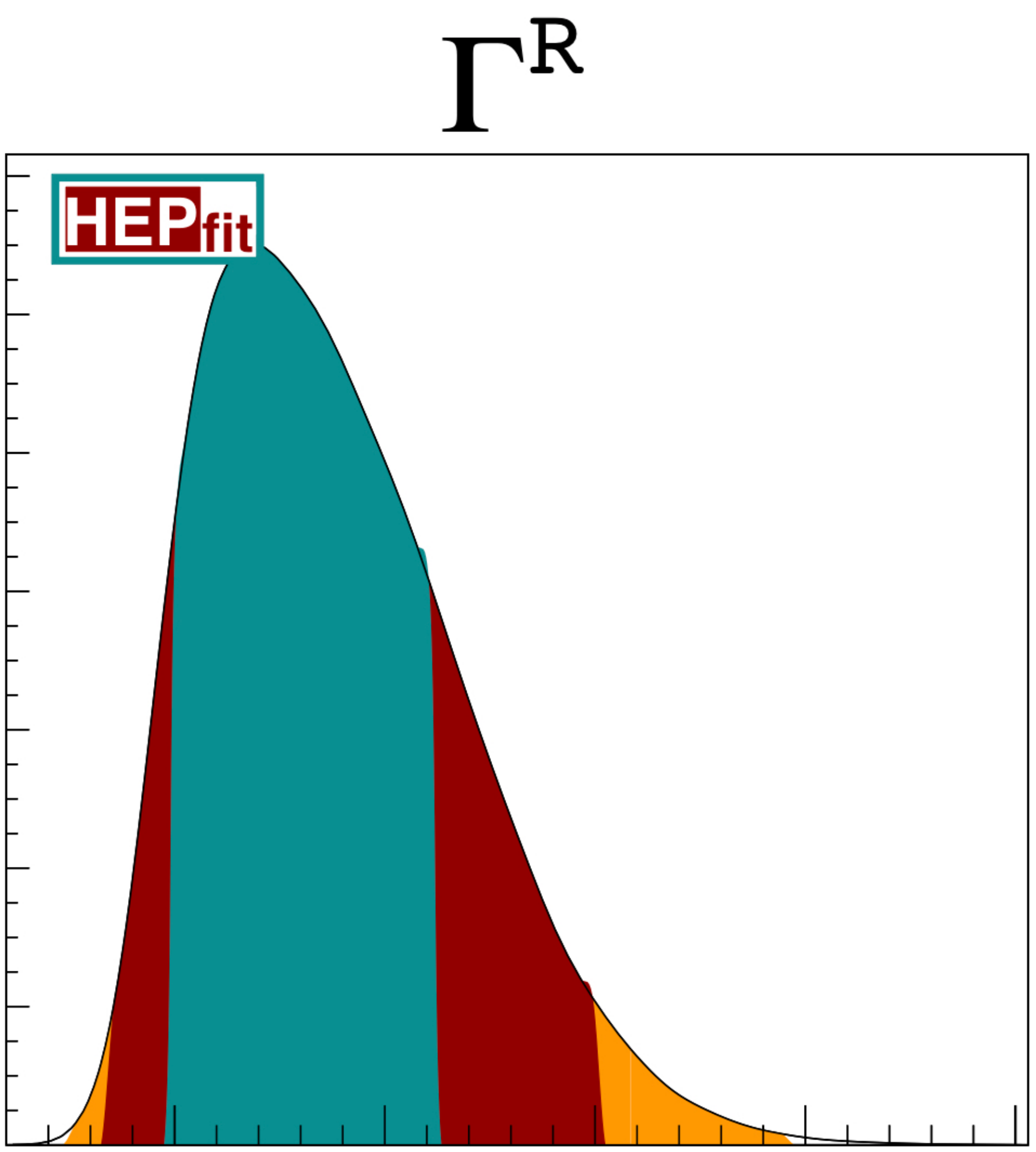}}
\vspace{-1.85em} \\
\subfigure{\includegraphics[width=.24\textwidth]{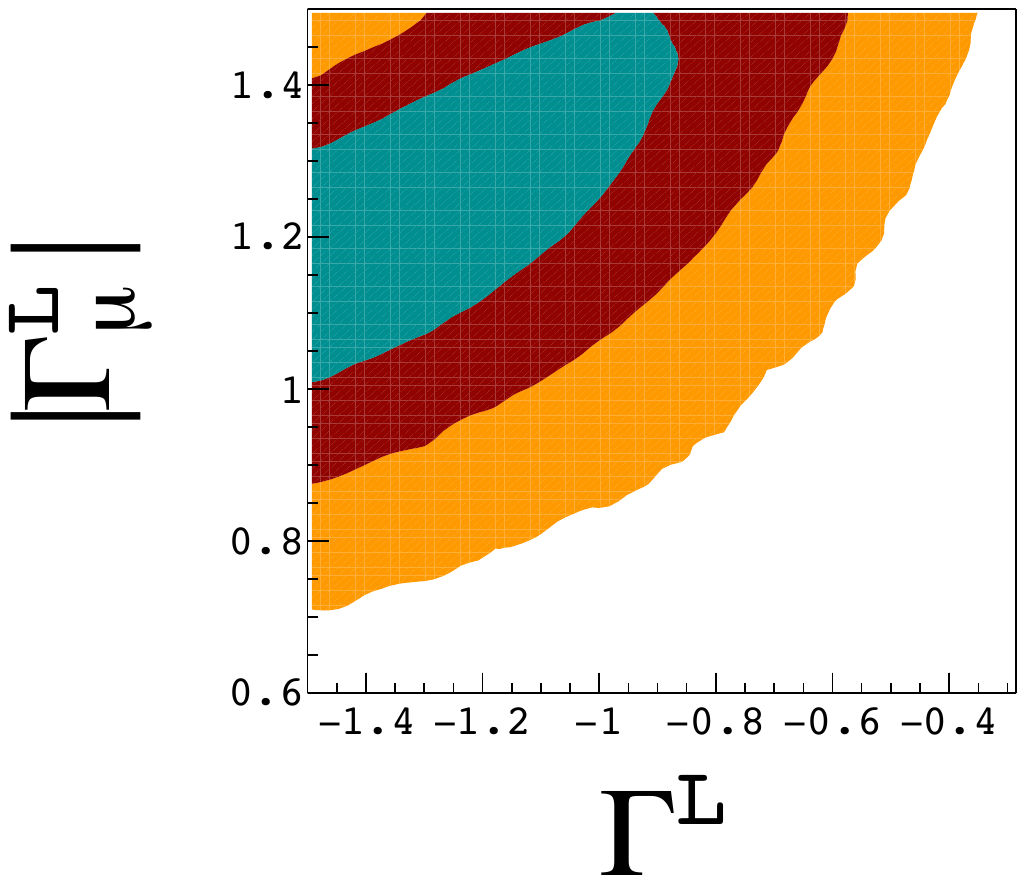}}
\subfigure{\includegraphics[width=.168\textwidth]{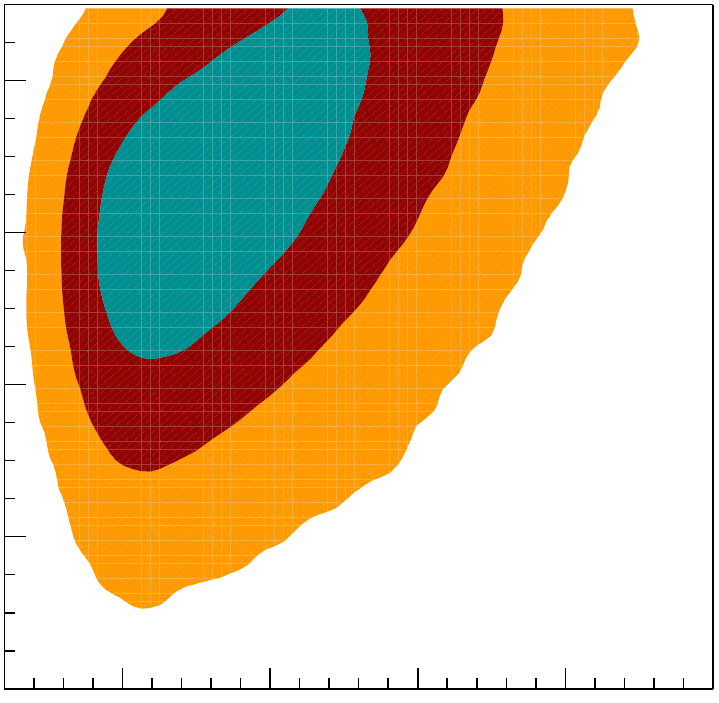}}
\subfigure{\includegraphics[width=.168\textwidth]{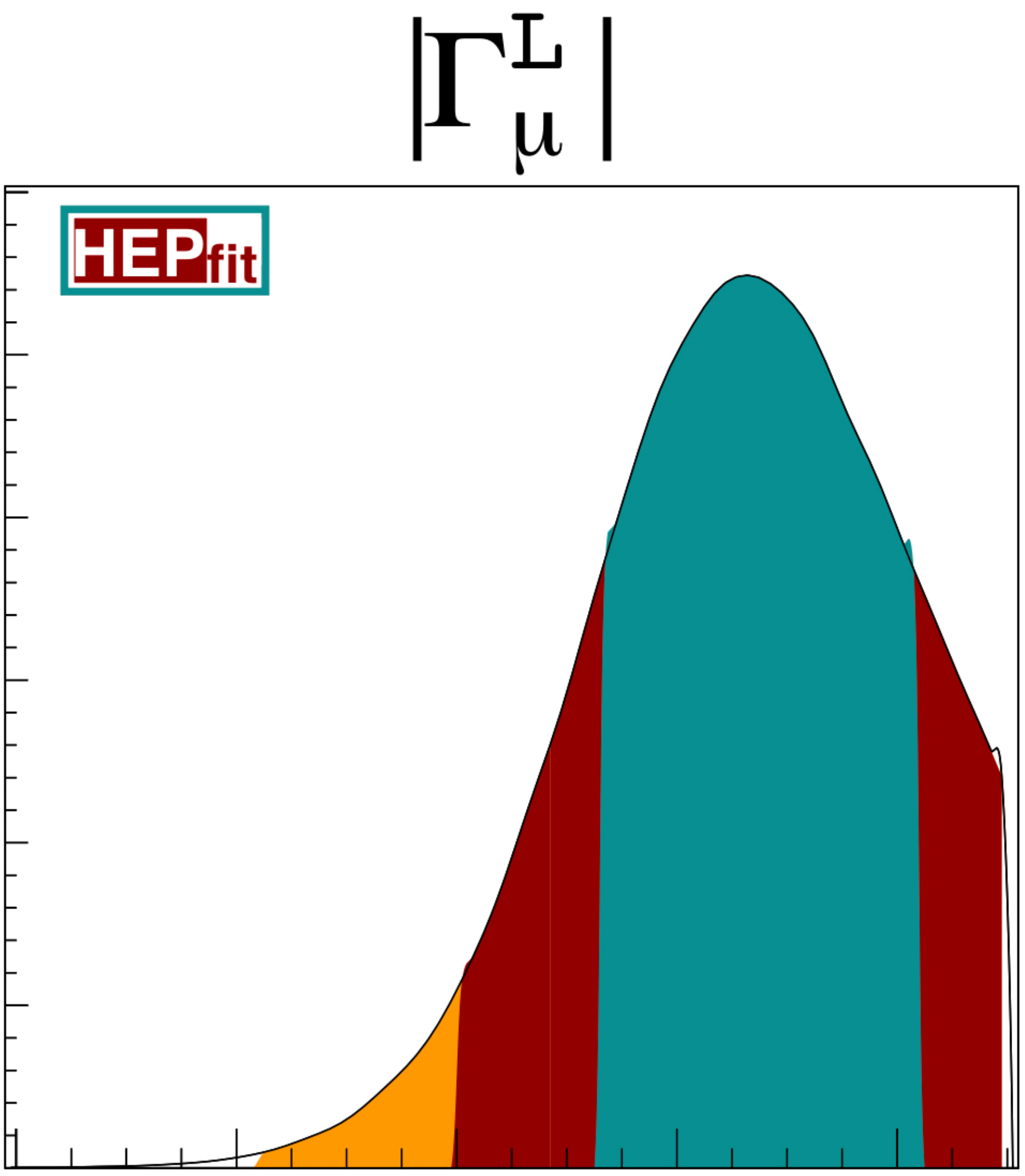}}
\vspace{-1.85em} \\
\subfigure{\includegraphics[width=.24\textwidth]{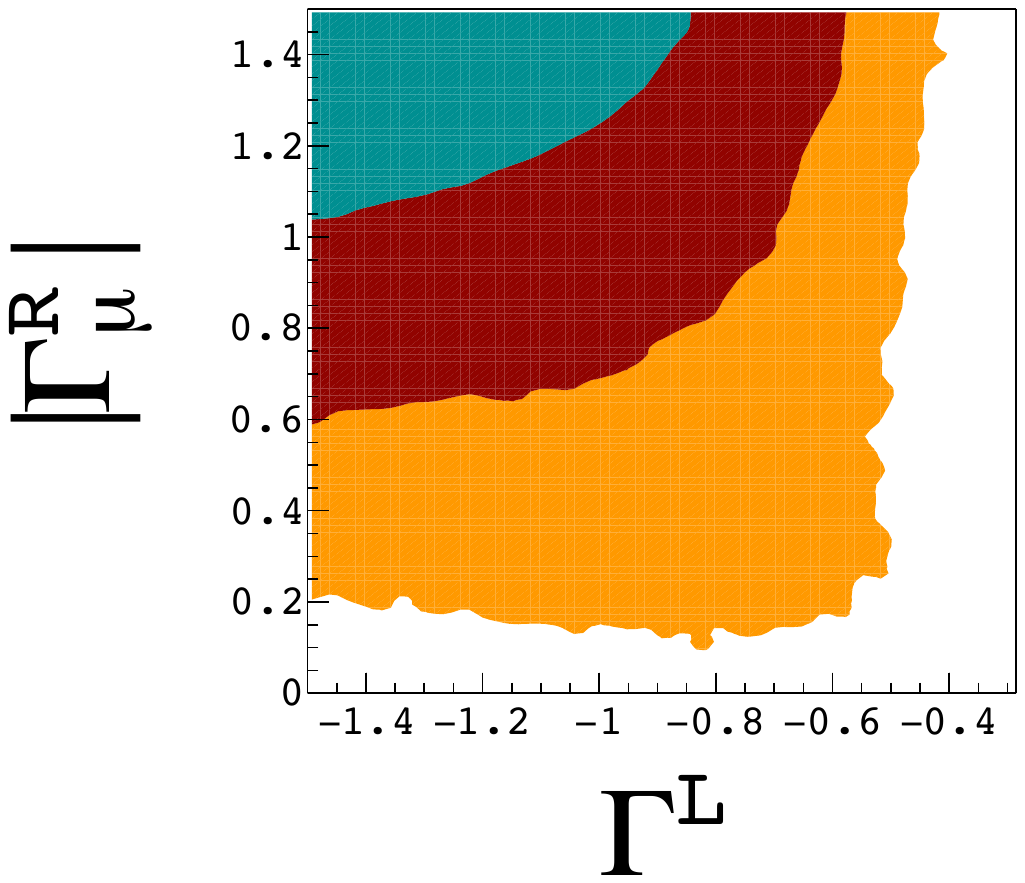}}
\subfigure{\includegraphics[width=.168\textwidth]{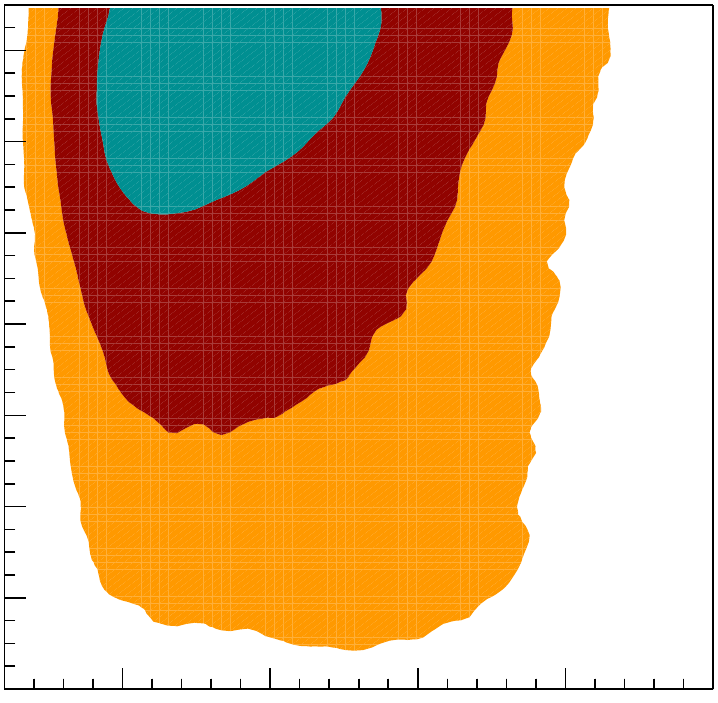}}
\subfigure{\includegraphics[width=.168\textwidth]{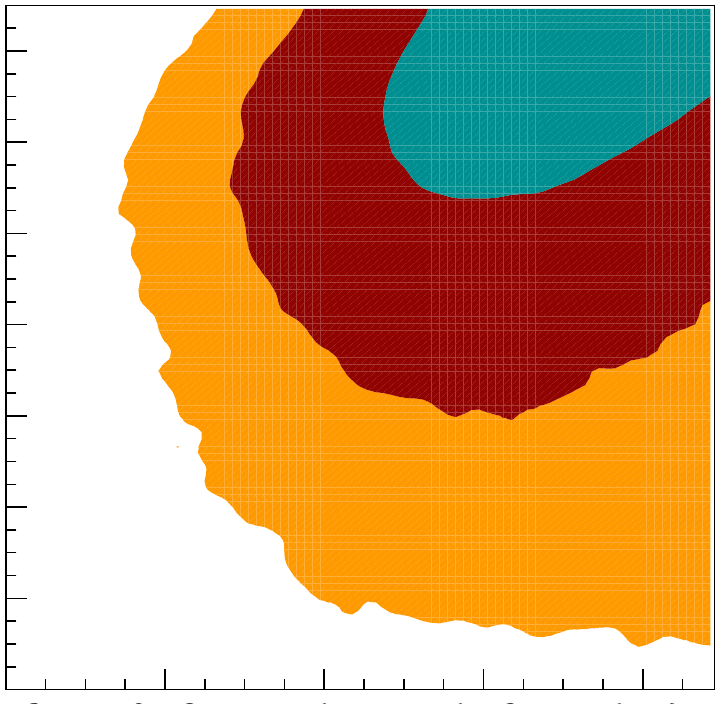}}
\subfigure{\includegraphics[width=.166\textwidth]{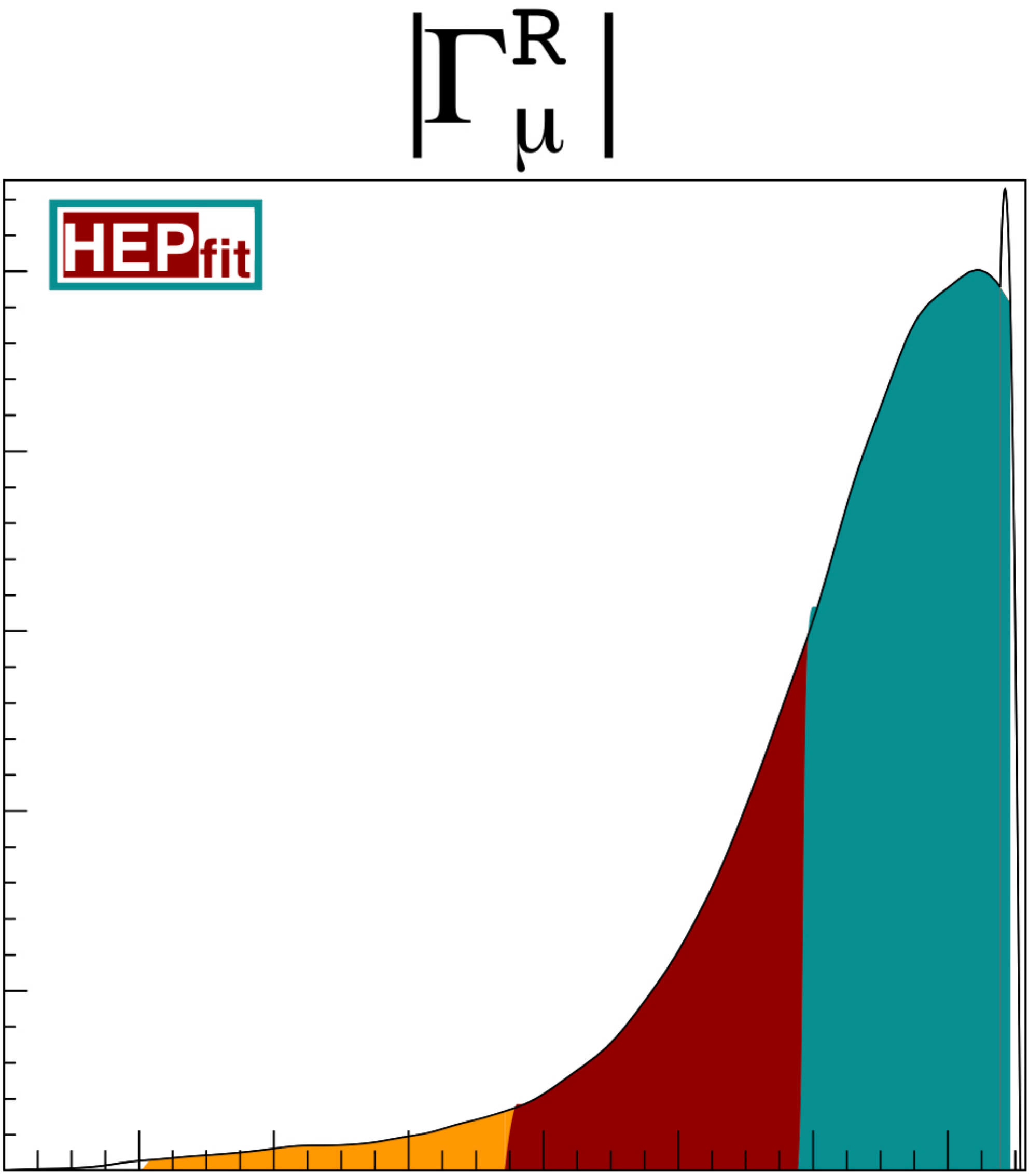}}
\vspace{-1.65em} \\
\subfigure{\includegraphics[width=.235\textwidth]{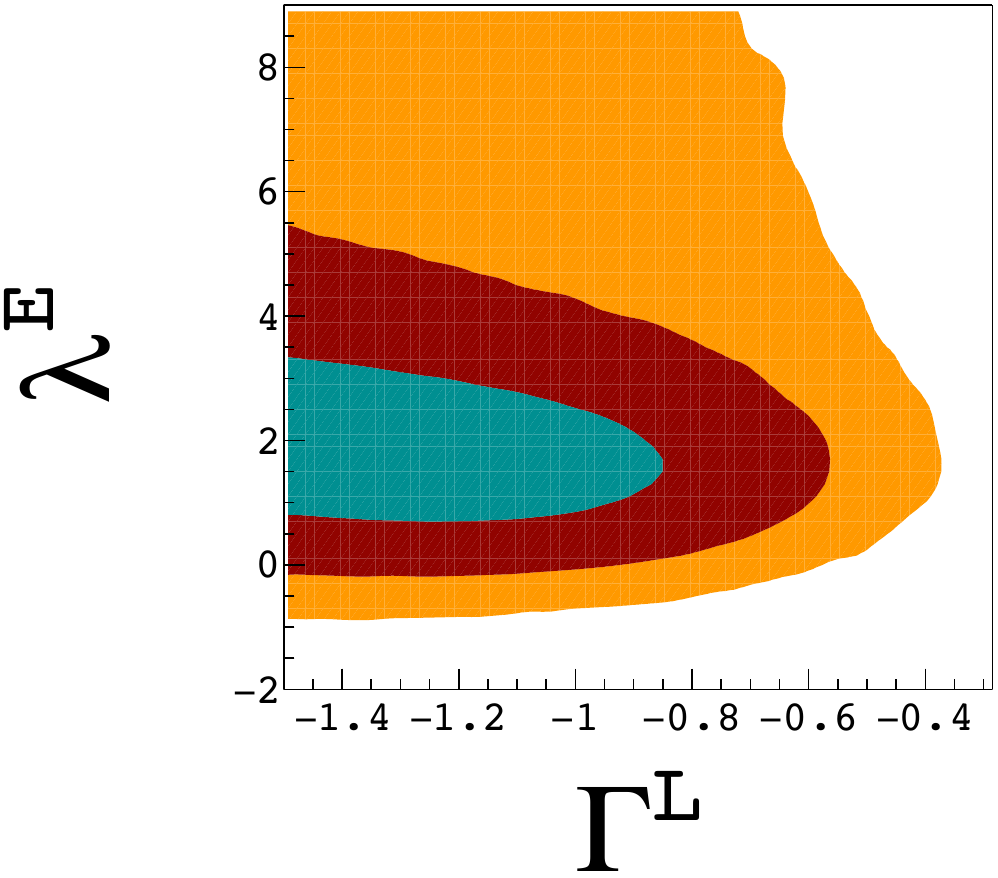}}
\subfigure{\includegraphics[width=.17\textwidth]{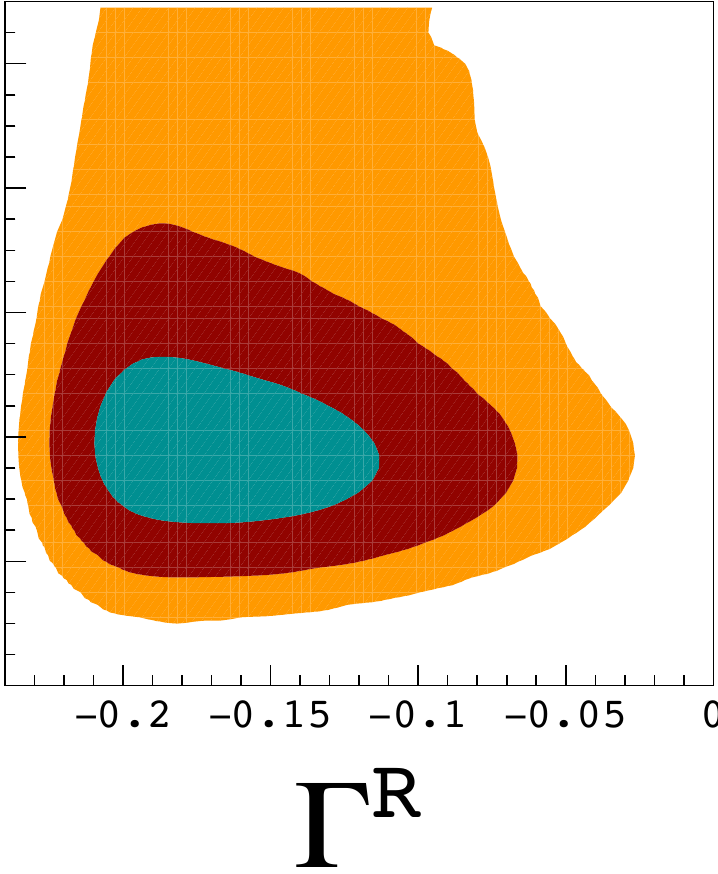}}
\subfigure{\includegraphics[width=.166\textwidth]{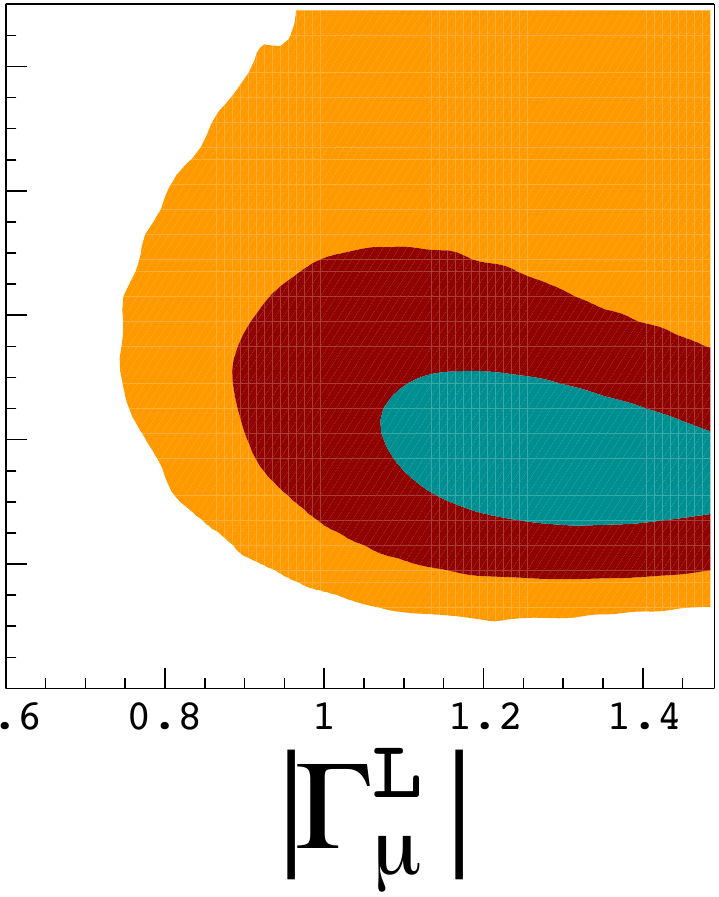}}
\subfigure{\includegraphics[width=.168\textwidth]{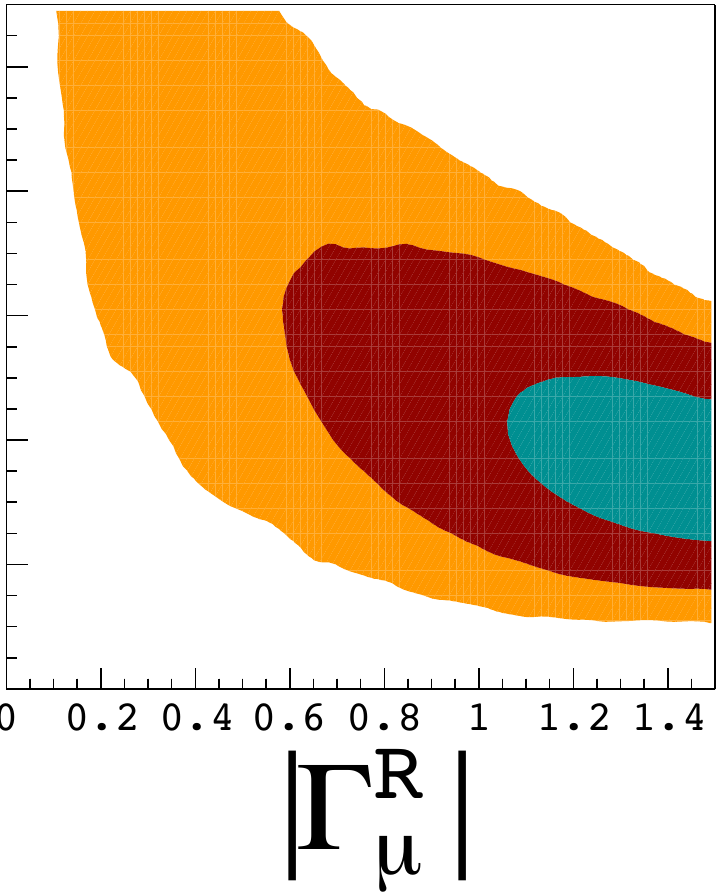}}
\subfigure{\includegraphics[width=.188\textwidth]{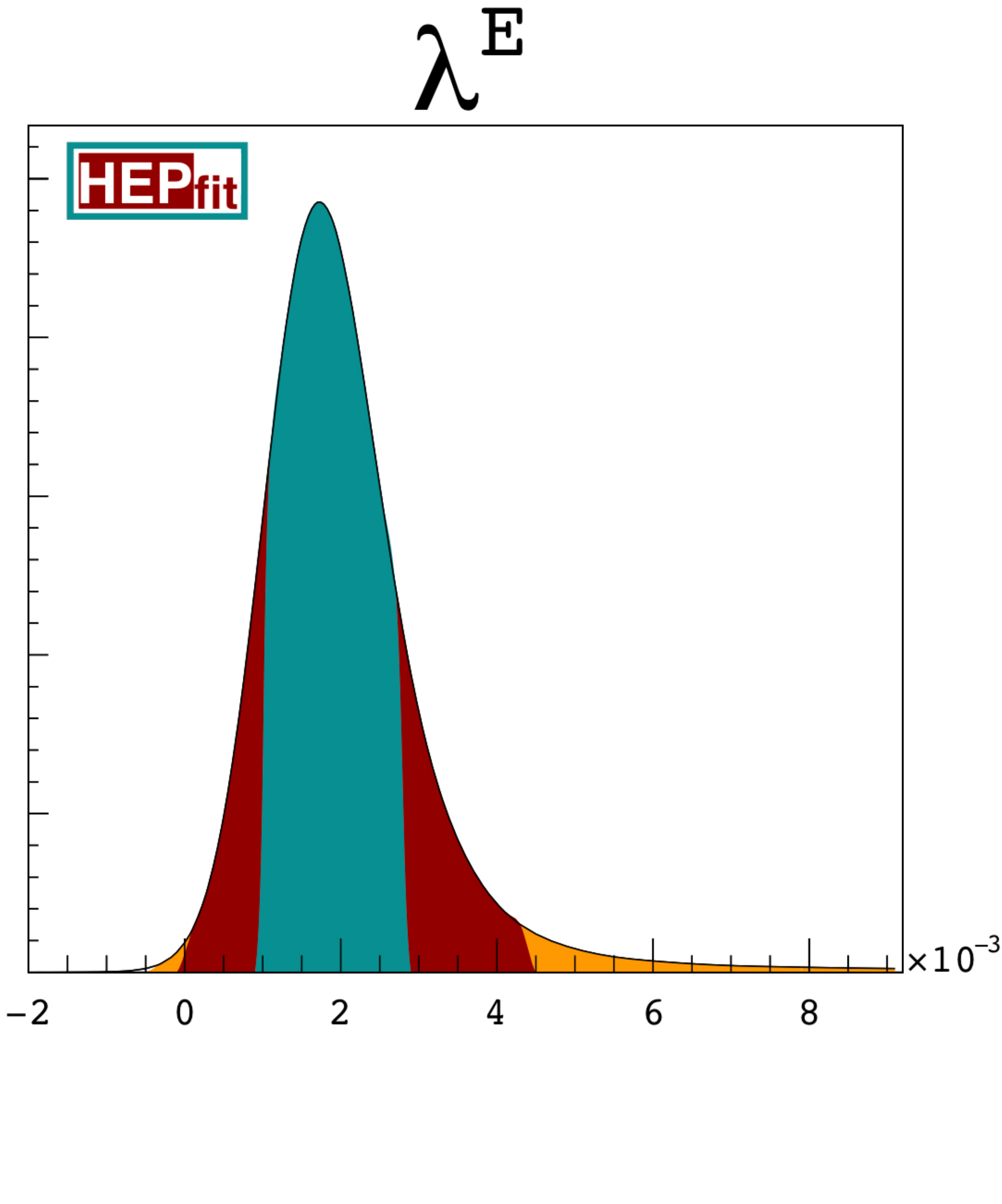}}
\caption{The $1D$ marginalized posterior distributions of the parameters from the fit described in sec.~\ref{sec:numerics}, together with the $2D$ correlations between them. The green, red and orange regions correspond to $68\%$, $95\%$ and $99\%$ probability regions, respectively.
\label{fig:posteriors}}
\end{figure}
Here we show the $1D$ marginalized posterior distributions of the parameters from the global fit described in Sec.~\ref{sec:numerics}, together with the $2D$ combined correlations between these parameters. The results are summarized in Fig.~\ref{fig:posteriors}. We recall that, for all couplings $\Gamma$, we imposed $|\Gamma| \leq 1.5$ such that perturbativity is satisfied. The consequences of such this choice are evident in the posterior distributions of $\Gamma^L$, $|\Gamma^L_\mu|$ and $|\Gamma^R_\mu|$, which are truncated because of this reason.

\footnotesize

\bibliographystyle{JHEP} 
\bibliography{BIB}

\end{document}